\documentclass[twocolumn]{aastex7}
\usepackage{enumerate}
\usepackage{amssymb, amsmath}
\usepackage{natbib}
\usepackage{color}
\usepackage{ulem}
\usepackage{appendix}
\usepackage{hyperref}
\usepackage[stable]{footmisc}
\usepackage{relsize}
\usepackage{mathtools}

\definecolor{red}{rgb}{1.0,0.0,0.0}

\def\surveyname{OASIS}

\begin{document}

\title{SCExAO/CHARIS and Gaia Direct Imaging and Astrometric Discovery of a Superjovian Planet\\ 3--4 $\lambda$/D from the Accelerating Star HIP 54515\footnote{Based in part on data collected at Subaru Telescope, which is operated by the National Astronomical Observatory of Japan.}}

\email{thayne.currie@utsa.edu}
\author[0000-0002-7405-3119]{Thayne Currie}
\altaffiliation{These authors contributed equally to this work as co-first authors.}
\affiliation{Department of Physics and Astronomy, University of Texas at San Antonio, San Antonio, TX 78249, USA}
 \affiliation{Subaru Telescope, National Astronomical Observatory of Japan, 
650 North A`oh$\bar{o}$k$\bar{u}$ Place, Hilo, HI  96720, USA}
\email[show]{thayne.currie@utsa.edu}
\author[0000-0002-6845-9702]{Yiting Li}
\altaffiliation{These authors contributed equally to this work as co-first authors.}
\affiliation{Department of Astronomy, University of Michigan, 1085 S. University, Ann Arbor, MI 48109, USA}
\email[show]{lyiting@umich.edu}
\author{Mona El Morsy}
\affiliation{Department of Physics and Astronomy, University of Texas at San Antonio, San Antonio, TX 78249, USA}
\email[]{mona.elmorsy@utsa.edu}
\author{Brianna Lacy}
\affiliation{Department of Astronomy and Astrophysics, University of California-Santa Cruz, Santa Cruz, CA, USA}
\email[]{blacy@ucsc.edu}
\author[orcid=0000-0001-5763-378X]{Maria Vincent}
\affiliation{Institute for Astronomy, University of Hawai'i, 640 N. Aohoku Pl., Hilo, HI 96720, USA}
\email[]{mariavin@hawaii.edu}
\author[0000-0001-8103-5499]{Taylor L. Tobin}
\affiliation{Department of Astronomy, University of Michigan, 1085 S. University, Ann Arbor, MI 48109, USA}
\email[]{tltobin@umich.edu}
\author[0000-0002-4677-9182]{Masayuki Kuzuhara}
\affiliation{Astrobiology Center, NINS, 2-21-1 Osawa, Mitaka, Tokyo 181-8588, Japan}
\affiliation{National Astronomical Observatory of Japan, 2-21-1 Osawa, Mitaka, Tokyo 181-8588, Japan}
\email[]{m.kuzuhara@nao.ac.jp}
\author[0000-0001-6305-7272]{Jeffrey Chilcote}
\affiliation{Department of Physics and Astronomy, University of Notre Dame,
Nieuwland Science Hall, Notre Dame, IN 46556, USA}
\email[]{jchilcot@nd.edu}
\author[]{Olivier Guyon}
 \affiliation{Subaru Telescope, National Astronomical Observatory of Japan,
650 North A`oh$\bar{o}$k$\bar{u}$ Place, Hilo, HI  96720, USA}
\affiliation{Astrobiology Center, 2-21-1 Osawa, Mitaka, Tokyo 181-8588, Japan}
\email[]{guyon@naoj.org}
\author{Ziying Gu}
\affiliation{Department of Astronomy, Faculty of Science, The Uniersity of Tokyo, 7-3-1 Hongo, Bunkyo-ku, Tokyo 113-0033, Japan}
\affiliation{Astrobiology Center, 2-21-1 Osawa, Mitaka, Tokyo 181-8588, Japan}
\email[]{guziying@g.ecc.u-tokyo.ac.jp}
\author{Danielle Bovie}
\affiliation{Department of Physics and Astronomy, University of Texas at San Antonio, San Antonio, TX 78249, USA}
\email[]{danielle.bovie@my.utsa.edu}
\author{Dillon Peng}
\affiliation{Department of Physics and Astronomy, University of Notre Dame,
Nieuwland Science Hall, Notre Dame, IN 46556, USA}
\email[]{dpeng@nd.edu}


\author[0000-0003-0115-547X]{Qier An}
\affiliation{Department of Physics and Astronomy Johns Hopkins University, Baltimore, MD 21218, USA}
\email[]{qan4@jh.edu}

\author{Timothy D Brandt}
\affiliation{Space Telescope Science Institute, Baltimore, MD, USA}
\email[]{tbrandt@stsci.edu }
\author{Vincent Deo}
 \affiliation{Subaru Telescope, National Astronomical Observatory of Japan,
650 North A`oh$\bar{o}$k$\bar{u}$ Place, Hilo, HI  96720, USA}
\email[]{vdeo@naoj.org}
\author{Robert de Rosa}
\email[]{rderosa@eso.org}
\affiliation{European Southern Observatory, Alonso de Córdova 3107, Vitacura, Santiago, Chile}
\author{Tyler D Groff}
\affiliation{NASA-Goddard Space Flight Center, Greenbelt, MD USA}
\email[]{tyler.d.groff@nasa.gov}
\author{Markus Janson}
\affiliation{Department of Astronomy, Stockholm University, AlbaNova University Center, Stockholm, 10691, Sweden}
\email[]{markus.janson@astro.su.se}
\author{N. J. Kasdin}
\affiliation{Department of Mechanical and Aerospace Engineering Princeton University, Princeton, New Jersey 08544, USA}
\email[]{jkasdin@princeton.edu}
\author[0000-0002-3047-1845]{Julien Lozi}
 \affiliation{Subaru Telescope, National Astronomical Observatory of Japan,
650 North A`oh$\bar{o}$k$\bar{u}$ Place, Hilo, HI  96720, USA}
\affiliation{Astrobiology Center, 2-21-1 Osawa, Mitaka, Tokyo 181-8588, Japan}
\email[]{lozi@naoj.org}
\author{Christian Marois}
\affiliation{National Research Council, Herzberg Astronomy and Astrophysics, Victoria, BC, Canada}
\email{christian.marois@nrc-cnrc.gc.ca}
\author{Bertrand Mennesson}
\affiliation{Jet Propulsion Laboratory, California Institute of Technology, Pasadena, California, United States}
\email[]{Bertrand.Mennesson@jpl.nasa.gov}
\author[0000-0001-6177-1333]{Naoshi Murakami}
\affiliation{Astrobiology Center, NINS, 2-21-1 Osawa, Mitaka, Tokyo 181-8588, Japan}
\affiliation{National Astronomical Observatory of Japan, 2-21-1 Osawa, Mitaka, Tokyo 181-8588, Japan}
\affiliation{The Graduate University for Advanced Studies, SOKENDAI, 2-21-1 Osawa, Mitaka, Tokyo 181-8588, Japan}
\affiliation{Faculty of Engineering, Hokkaido University, Kita 13, Nishi 8, Kita-ku, Sapporo, Hokkaido 060-8628, Japan}
\email[]{naoshi.murakami@nao.ac.jp}

\author{Eric Nielsen}
\affiliation{Department of Astronomy, New Mexico State University, P. O. Box 30001, MSC 4500, Las Cruces, NM 88003, USA}
\email[]{nielsen@nmsu.edu}

\author{Sabina Sagynbayeva}
\affiliation{ Department of Physics and Astronomy, Stony Brook University, Stony Brook, NY 11794, USA  }
\email[]{sabina.sagynbayeva@stonybrook.edu}
\author[0000-0002-9372-5056]{Nour Skaf}
\affiliation{Institute for Astronomy, University of Hawai’i at Manoa, Hilo, HI 96720-2700, US}
\email[]{nskaf@hawaii.edu}

\author{William Thompson}
\affiliation{National Research Council, Herzberg Astronomy and Astrophysics, Victoria, BC, Canada}
\email[]{William.Thompson@nrc-cnrc.gc.ca}
\author[0000-0002-6510-0681]{Motohide Tamura}
\affiliation{Astrobiology Center, 2-21-1 Osawa, Mitaka, Tokyo 181-8588, Japan}
\affiliation{National Astronomical Observatory of Japan, 2-21-1 Osawa, Mitaka, Tokyo 181-8588, Japan}
\affiliation{Department of Astronomy, Faculty of Science, The University of Tokyo, 7-3-1 Hongo, Bunkyo-ku, Tokyo 113-0033, Japan}
\email[]{motohide.tamura@astron.s.u-tokyo.ac.jp}

\author[0000-0002-6879-3030]{Taichi Uyama}
\affiliation{Department of Physics and Astronomy, California State University Northridge, 18111 Nordhoff Street, Northridge, CA 91330, USA}
\email[]{taichi.uyama.astro@gmail.com}

\author{Sebastien Vievard}
 \affiliation{Subaru Telescope, National Astronomical Observatory of Japan,
650 North A`oh$\bar{o}$k$\bar{u}$ Place, Hilo, HI  96720, USA}
\email[]{vievard@naoj.org}

\author[0000-0002-5903-8316]{Alice Zurlo}
\affiliation{Instituto de Estudios Astrof\'isicos, Facultad de Ingenier\'ia y Ciencias, Universidad Diego Portales, Av. Ej\'ercito Libertador 441, Santiago, Chile}
\affiliation{Millennium Nucleus on Young Exoplanets and their Moons (YEMS)}
\email[]{alice.zurlo@mail.udp.cl}

\shortauthors{Currie and Li et al. 2025}
\begin{abstract}
We present the discovery of a superjovian planet around the young A5 star HIP 54515, detected using precision astrometry from the Hipparcos Gaia Catalogue of AccelerationsVand high-contrast imaging with SCExAO/CHARIS from the recently commenced \surveyname\ program.   SCExAO/CHARIS detects HIP 54515 b in five epochs 0\farcs{}145--0\farcs{}192 from the star ($\sim$3--4 $\lambda$/D at 1.65 $\mu m$), exhibiting clockwise orbital motion.  
HIP 54515 b lies near the M/L transition with a luminosity of log(L/L$_{\rm \odot}$) $\sim$ -3.52 $\pm$ 0.03. 
Dynamical modeling constrains its mass and mass ratio to be ${17.7}_{-4.9}^{+7.6}$ $M_{\rm Jup}$ and ${0.0090}_{-0.0024}^{+0.0036}$
and favors a $\sim$25 au semimajor axis.  HIP 54515 b adds to a growing list of superjovian planets with moderate eccentricities ($e$ $\approx$ 0.4).  Now the third planet discovered from surveys combining high-contrast extreme adaptive optics imaging with precision astrometry, HIP 54515 b should help improve empirical constraints on the luminosity evolution and eccentricity distribution of the most massive planets. It may also provide a key technical test of the Roman Space Telescope Coronagraph Instrument's performance in the low stellar flux, small angular separation limit and a demonstration of its ability to yield constrainable planet spectral properties.

\end{abstract}

\section{Introduction}

Over the past two decades, ground-based telescopes equipped with conventional and now \textit{extreme} adaptive optics (AO) systems have revealed direct images of jovian planets orbiting $\sim$10--100 au from nearly two-dozen nearby, young stars \citep[see ][and references therein]{Currie2023b}.    
However, results from large-scale, `blind' direct imaging surveys responsible for most of these discoveries suggest that wide-separation giant planets and brown dwarfs imageable with current facilities are rare (e.g., \citealt{Nielsen2019, Vigan2015}). 

More recently, direct imaging efforts have shifted toward targeted searches of stars showing dynamical evidence for companions based on the precision astrometry \citep{Brandt2021,Kervella2022}.   
While such astrometry-selected surveys are still in their infancy, they have already yielded the discovery of two planets -- HIP 99770 b and AF Lep b \citep{Currie2023a,Franson_2023, deRosa2023,Mesa2023}.  Additionally, they have revealed one companion that is either a massive planet or low-mass brown dwarf (HIP 39017 b; \citealt{Tobin2024}),  about half a dozen brown dwarf companions \citep{Currie2020a,Bonavita2022,Kuzuhara2022,Li2023,Li2024}\footnote{Planets may also be identifiable by comparing Gaia accelerations to resolved structures in disks suggestive of massive, forming companions \citep{Ribas2025}.}. Early analysis suggests that they may yield a higher detection rate of young planets and brown dwarfs \citep{ElMorsy2024a}. 

Furthermore, the synergy between direct imaging and absolute astrometry provides a uniquely powerful framework for characterizing planetary systems. Relative astrometry from high-contrast imaging, when jointly modeled with proper motion accelerations from the Hipparcos and Gaia, directly constrains companion dynamical masses without relying on luminosity evolution models \citep{Brandt2019}, which are sensitive to assumptions about age and composition \citep{Moro_2010}.   For systems with a particularly high signal-to-noise ratio (SNR) acceleration, companion masses and orbital parameters can be constrained to within several percent precision from long-term monitoring \citep[e.g.][]{Brandt2019,Balmer2025,ElMorsy2024b}. 

Many companions discovered from accelerating star surveys can be well suited for atmospheric characterization at other wavelengths with JWST and the Roman Coronagraph Instrument \citep[]{ElMorsy2024b,ElMorsy2025,Bovie2025}.  Follow-up higher-resolution spectra then tie atmospheric properties like clouds and chemistry to objects of a given mass and age \citep{Bovie2025,ElMorsy2024b}.   
Motivated by these advantages, we initiated an intensive direct imaging program systematically targeting accelerating stars: the Observing Accelerators with SCExAO Imaging Survey  \citep[OASIS; PI: T. Currie, Co-PI: M. Kuzuhara][]{ElMorsy2024a}.   

In this paper, we present the first discovery of an exoplanet from OASIS: a superjovian companion located 0\farcs{}145--0\farcs{}192 ($\sim$3--4 $\lambda$/D at 1.65 $\mu m$) from the young mid-A star HIP~54515. An overview of the host star HIP~54515 is provided in Section~\ref{sec:system_properties}. Observations and the direct detection of the companion are described in Section~\ref{sec:data}. Our spectrophotometric and orbital analyses are presented in Section~\ref{sec:spec_analysis} and Section~\ref{sec:astrometric_analysis}, respectively. Finally, further discussion is included in Section~\ref{sec:discussion}, including an assessment of HIP~54515~b’s suitability as a target for the Roman Coronagraph Instrument (CGI).

\begin{figure*}
\centering   \includegraphics[width=0.46\textwidth]{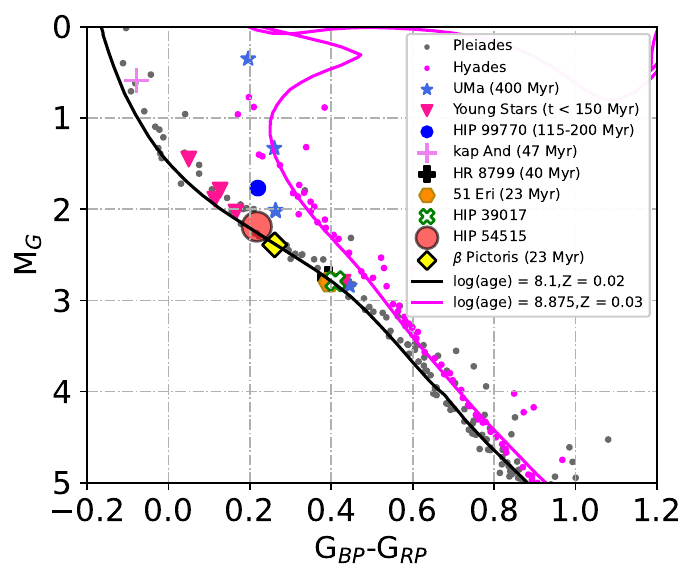}
   \includegraphics[width=0.5\textwidth]{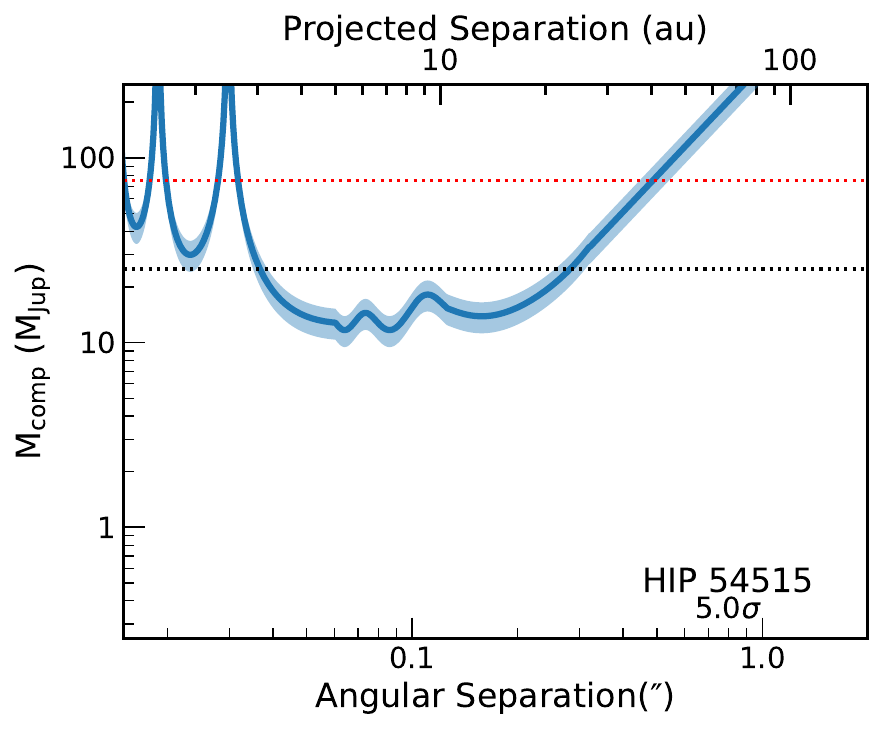}
   \vspace{-.12in}
   \caption{(left) The Gaia color-magnitude diagram for the Pleiades (gray) and Hyades (dark red) compared to PARSEC isochrones for 112 Myr and 750 Myr (roughly equal to the clusters' estimated ages), young stars with interferometrically measured ages $\lesssim$ 150 Myr, stars in the 400 Myr old Ursa Majoris Moving group, and other young stars with imaged substellar companions \citep{Jones2015,Jones2016,Jones2016PhD,Currie2023a,Tobin2024}. (right) Minimum mass vs. projected separation needed to explain HIP 54515's proper motion anomaly, following the approach of \citet{Kervella2022} with the stellar/substellar boundary overplotted as a red dotted line and approximate brown dwarf/planet boundary overplotted as a black dotted line.  The shaded region denotes the minimum mass 68\% confidence interval derived from uncertainties in the stellar mass and the proper motion.}
   \label{fig:cmd} 
\end{figure*}
\section{System Properties}
\label{sec:system_properties}
HIP~54515 (HD~96855, 66~Leo) is an A-type star located at a distance of 83.2~pc \citep{GAIA2023}.  Literature spectral classifications range from A2 to A5, with some disagreement about whether it is a dwarf or a (sub)giant \citep{Cannon1993,Pickles2010}.  The most recent literature estimate favors A5V \citep{Pickles2010}.  An A5V classification is consistent with the star's optical properties: its $(B-V)_0$ color (0.165) and absolute V band magnitude (2.21) are nearly indistinguishable  from the high-$v\sin i$ A5V standard star HD~23886.  Adopting the spectral type/effective temperature/mass scale from \citep{Pecaut2013} and assuming an uncertainty of two subtypes, HIP 54515 A's mass is 1.88 $\pm$ 0.11 $M_{\rm \odot}$ and temperature is 8100$^{+500}_{-340}$ $K$.

Literature age estimates for the star vary wildly, from 50.1 $\pm$ 39.1 Myr \citep{Tetzlaff2011} to a $\sim$500 Myr Gaia-DR3 Flames estimate \citep{GAIA2023}.  
Its position in a \textit{Gaia} color-magnitude diagram \citep[e.g.][]{ElMorsy2024b} is similar to the $\sim$23 Myr-old A6V star, $\beta$ Pictoris \citep{Mamajek2014}, which likewise lies very close to the Pleiades sequence and other stars with interferometrically-determined ages of less than 150 Myr \citep{Jones2016PhD} (Figure \ref{fig:cmd}, left panel).  It lies closer to the Pleiades sequence than the A5V star HIP 99770 (115-200 Myr) and far closer than Ursa Majoris members: its position is inconsistent with 500 Myr solar-metallicity isochrones.   A Bayesian age estimate (Gu, Kuzuhara et al. in preparation) using the PARSEC isochrones and uninformative priors on metallicity favors older ages (400 $\pm$ 250 Myr), but imposing a solar metallicity recovers near-Pleiades ages ($\approx$ 200 Myr).  Its kinematics $(U, V, W) = (-22.7, -17.6, -0.7)\,\mathrm{km\,s}^{-1}$ are broadly similar to members of the 50 Myr old Carina-Near moving group, though the crowded UVW phase space makes any firm association unlikely, and Banyan-$\Sigma$ favors a field object identification \citep{Gagne2018}.  No infrared excess has been detected from Wide-field Infrared Survey Explorer observations, and no debris disk is known.  Based on its similarity to Pleiades members and $\beta$ Pic, and relative youth compared to HIP 99770, we adopt an age estimate of 115$^{+85}_{-92}$ Myr.

The Hipparcos Gaia Catalogue of Accelerations (HGCA) reveals that HIP 54515 has a $\sim$5-$\sigma$ significant deviation from linear motion across the sky.  Its low renormalized unit weight error (RUWE $\sim$ 1.025) disfavors the presence of a massive, unresolvable short-period companion.   Following the methods described in \citet{Kervella2022}, we estimate that HIP 54515's acceleration could be caused by a companion with a mass of $\approx$10--20 $M_{\rm Jup}$ at 0\farcs{}1--0\farcs{}3 (Figure \ref{fig:cmd}, right panel).  A thorough search of archival ground-based data shows that HIP 54515 has never been targeted for a planet search with any technique.  Thus, we targeted this star as a part of OASIS to directly image planets around \textit{accelerating stars}.

\begin{deluxetable*}{lllllllll}[ht]
     \tablewidth{0pt}
    \tablecaption{HIP 54515 Observing Log\label{obslog}}
    \tablehead{\colhead{UT Date} & \colhead{Instrument} &  \colhead{Seeing$^{c}$ (\arcsec{})} &{Filter} & \colhead{$\lambda$ ($\mu m$)$^{a}$} 
    & \colhead{$t_{\rm exp}$} & \colhead{$N_{\rm exp}$} & \colhead{$\Delta$PA ($^{o}$)} & SNR (HIP 54515 b)}
    \startdata
    20221231 & AO188+SCExAO/CHARIS &  -$^{a}$ & $JHK$ & 1.16--2.37& 30.98 & 71 & 31.7  & 16.2 (13.0)$^{c}$\\
    20240220 & AO188+SCExAO/CHARIS &  0.6 & $JHK$ & 1.16--2.37& 25.08 & 146 & 62.0  & 9.3\\
    20240427 & AO188+SCExAO/CHARIS &  -$^{b}$ & $JHK$ & 1.16--2.37& 60.49 & 45 & 29.0  & 5.4\\
    20250314 & AO3K+SCExAO/CHARIS &  $\sim$1.5--2 & $JHK$ & 1.16--2.37& 30.98 & 182 & 32.8  & 7.0\\
    20250514 & AO3K+SCExAO/CHARIS &  0.6--0.8 & $JHK$ & 1.16--2.37& 30.98 & 199 & 65.0  & 11.0\\
    \enddata
   a) No seeing measurement recorded.  Qualitatively, conditions were excellent and characteristic of performance with seeing of $\lesssim$0\farcs{}4. 
   b) No seeing measurement recorded but AO correction was poorer than for other epochs.
   c) The parentheses note the SNR for the reduction used for analysis, which had a significantly less severe spectrophotometric and astrometric bias.
    \label{obslog_hip54515}
    \end{deluxetable*}

\section{Data}
\label{sec:data}
\subsection{Observations}
We observed HIP 54515 with SCExAO/CHARIS in its low spectral resolution (broadband) mode  covering the $JHK$ passbands simultaneously (1.16-2.37 $\mu$m at $\mathcal{R}$ $\sim$ 18) in five epochs between 30 December 2022 and 13 May 2025 as a part of the OASIS survey \citep{Currie2020b,ElMorsy2024a} (Table \ref{obslog_hip54515}).  For the 2024 epochs, SCExAO provided a sharper, second-order correction behind Subaru's facility AO system, AO188 \citep{Minowa2010}.  For the 2025 epochs, we conducted observations using the AO3K coupled with near-IR wavefront sensing \citep{Lozi2022}, an upgrade over AO188.  

We acquired all data in pupil tracking mode to enable angular differential imaging \citep[ADI;][]{Marois2006}.  Total exposure times ranged between 36.7 minutes and 102 minutes; parallactic angle rotations ranged between 31.7$^{o}$ and 65$^{o}$.  For all data, we used a Lyot coronagraph with a 0\farcs{}113-radius occulting spot.     We modulated the SCExAO deformable mirror to generate satellite spots used for astrometric and spectrophotometric calibration \citep[25 nm or 50 nm modulation amplitude][]{Jovanovic2015-astrogrids}.  

Observing conditions and AO performance varied significantly between epochs.   The first epoch had by far the best seeing conditions, resulting in the best image quality and a 1.6 $\mu m$ raw contrast at 0\farcs{}1--0\farcs{}25 about a factor of 3--10 deeper than other epochs.  Seeing during the March 2025 epoch was very poor -- up to 2$\arcsec{}$ with 30 $mph$ winds.   The 2025 March and May data were compromised by periodic low-wind effect \citep{Milli2018}; the May raw contrast was further degraded outside the coronagraph mask edge by a small cluster of stuck actuators on AO3K, which have since been fixed.  

\begin{figure*}[ht!]
 \includegraphics[width=0.333\textwidth,clip]{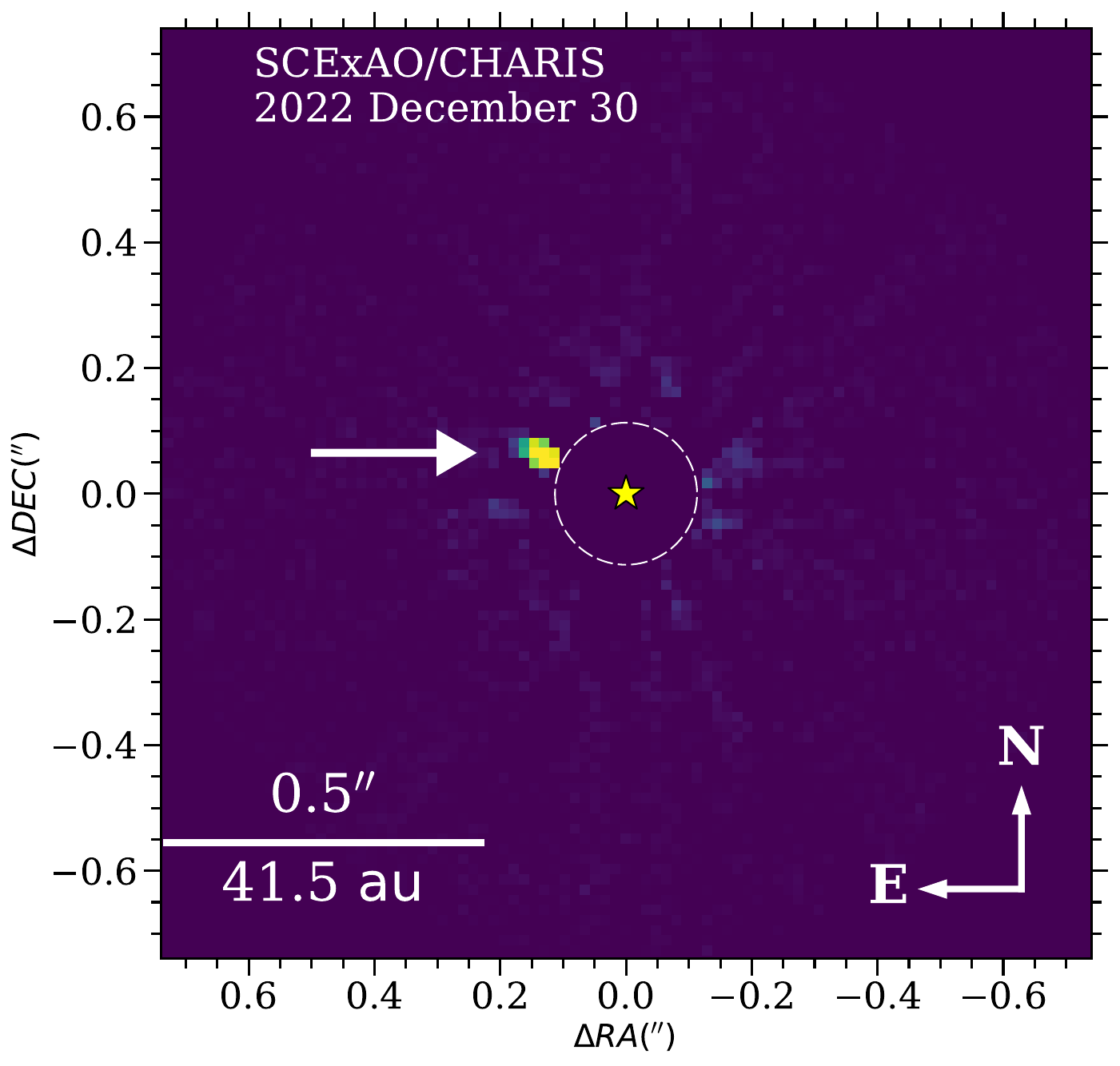}
   \includegraphics[width=0.333\textwidth,clip]{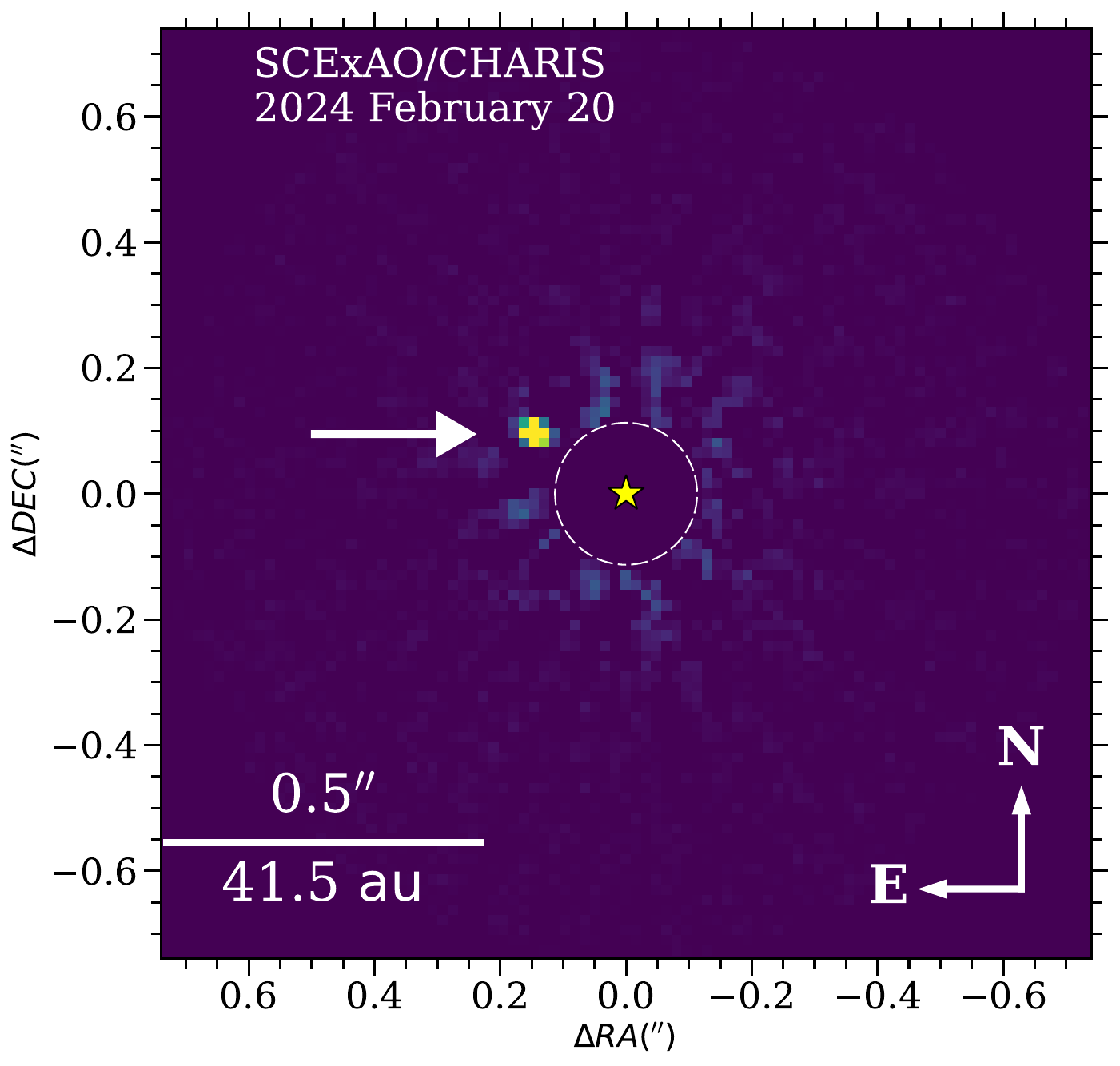}
   \includegraphics[width=0.333\textwidth,clip]{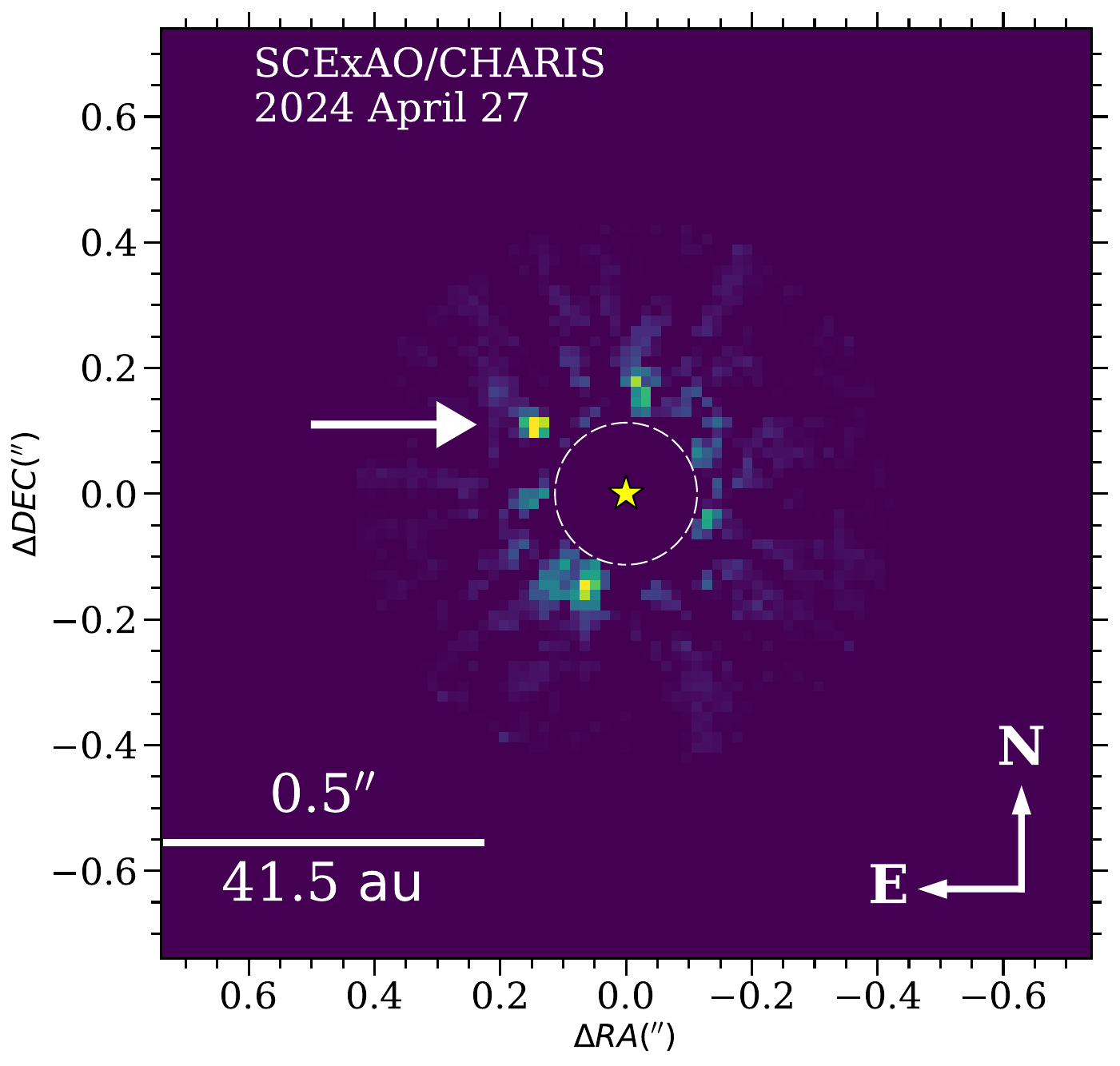}
   \includegraphics[width=0.333\textwidth,clip]{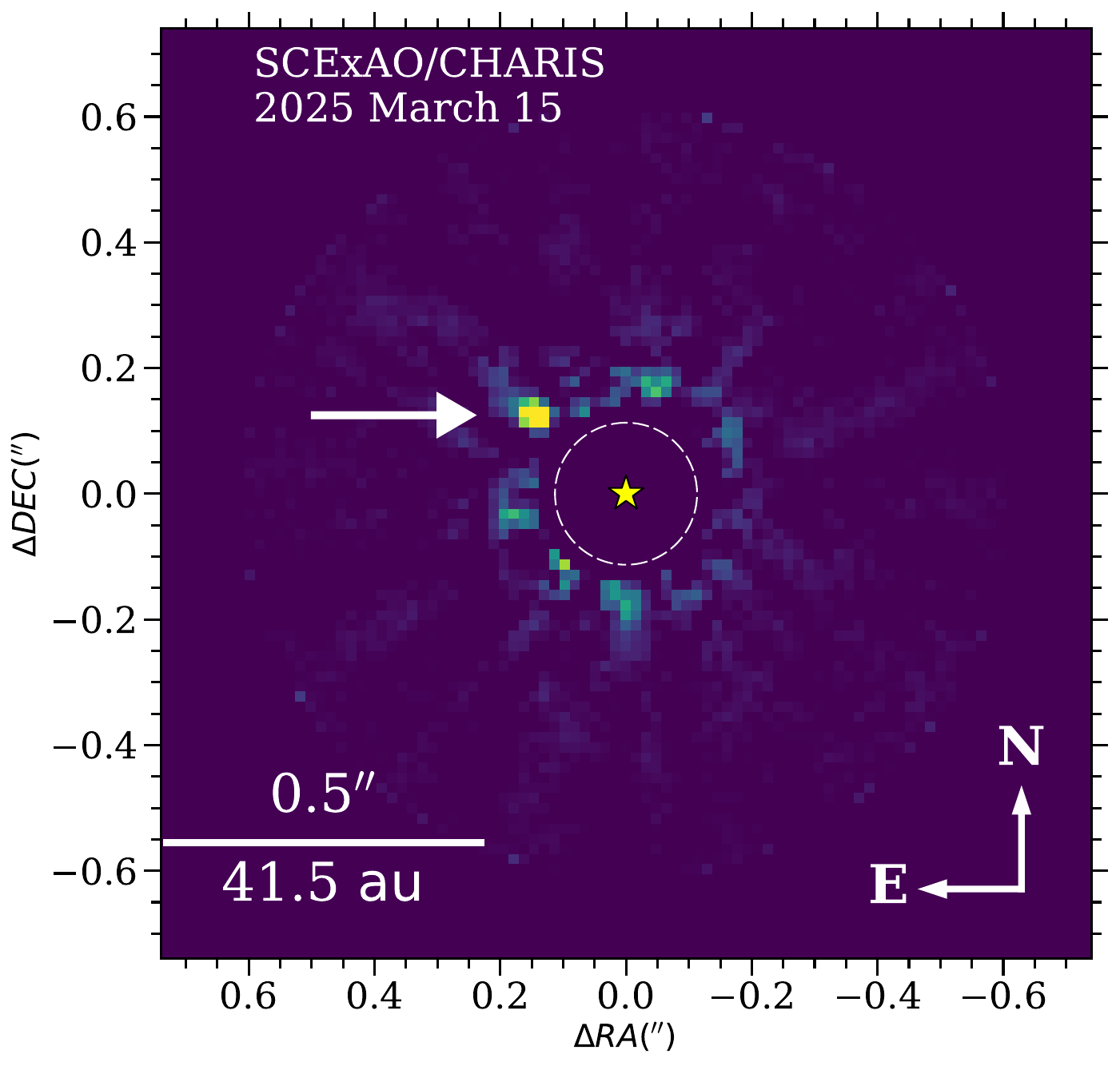}
   \includegraphics[width=0.333\textwidth,clip]{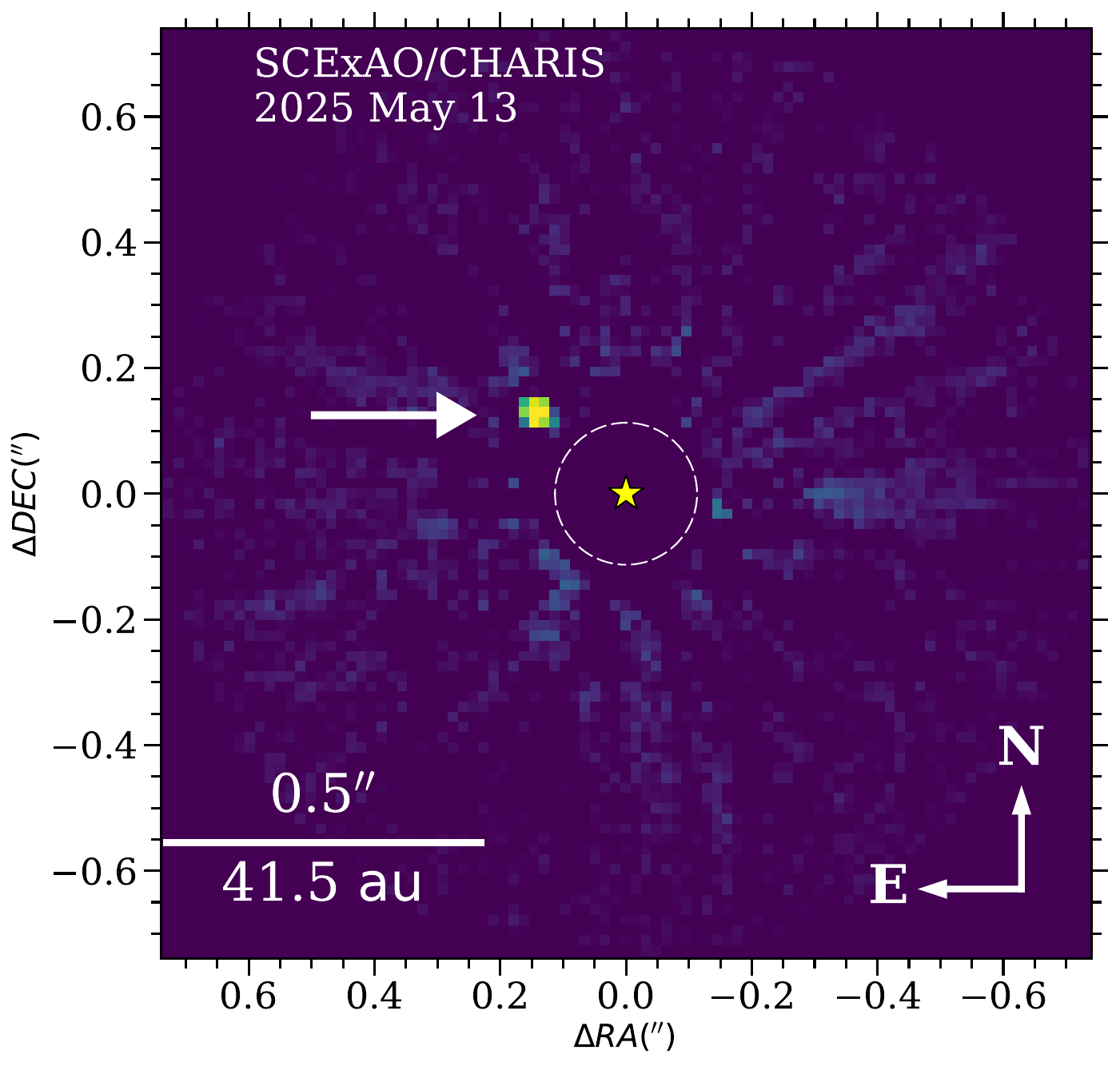}
  \caption{
Wavelength-collapsed CHARIS images over five epochs showing detections of HIP 54515 b.  The color scale is adjusted to saturate the signal within a FWHM-sized region, except for the April 2024 epoch where the scaling is adjusted to better differentiate the planet from residual speckle noise.   The companion PSF is exterior to the coronagraph mask edge in all epochs (dashed circle).
  }
  \label{fig:images}
\end{figure*}
\subsection{Data Reduction}
Following previous studies of accelerating stars \citep{Currie2020a,Kuzuhara2022,Currie2023b,Tobin2024}, we first extracted the CHARIS data cubes from the raw reads \citep{Brandt2017} \footnote{\url{https://github.com/PrincetonUniversity/charis-dep}}.  Then, we processed the data using the CHARIS Data Processing Pipeline \citep{Currie2020a}\footnote{\url{https://github.com/thaynecurrie/charis-dpp}}.   We subtracted sky frames from our data cubes to remove thermal emission and then registered each cube to a common center using a polynomial fit to x and y positions of each slice.   For spectrophotometric calibration, we performed photometry on the satellite spots and adopted a Kurucz stellar atmosphere model appropriate for an A5V star matched to HIP 54515 A's near-IR photometry.   Finally, we spatially filtered each cube slice: a simple radial profile subtraction for the 2024 February and 2025 May data and a moving-box median filter of 15 pixels for the other data sets.  As in prior work \citep{Bovie2025}, we removed cubes from analysis with degraded raw contrasts due to poorer AO corrections: the $N_{\rm exp}$ column in Table \ref{obslog_hip54515} lists the number of frames remaining.

For point-spread function (PSF) subtraction, we used the Adaptive, Locally Optimized Combination of Images algorithm in combination with ADI \citep[A-LOCI][]{Currie2012,Currie2015}. We explored the A-LOCI free parameter space, varying the rotation gap criterion ($\delta$), the optimization area (i.e. the area over which a reference PSF is computed; $N_{A}$), and the singular value decomposition cutoff ($SVD_{\rm lim}$) \citep[see also][]{Lafreniere2007,Marois2010b,Currie2012}. We also varied the number of frames used to build the reference PSF ($n_{\rm ref}$) via a correlation-based frame selection and explored the option to mask the subtraction region when computing the reference PSF \citep{Currie2015,Currie2018} 
\footnote{The detection of HIP 54515 b is robust at the $>$5-$\sigma$ level for the 2022 December, 2024 February, and 2025 May data sets for a wide range of parameter space, while a 5-$\sigma$ detection in the 2024 April and 2025 March data sets required more careful algorithm setups.  HIP 54515 b lies at a very small angular separation, and the star undergoes slow parallactic angle changes since it does not transit near zenith.  Thus, the parameter space yielding the best point source sensitivity required a small rotation gap (typically $\delta$ $\sim$ 0.3-0.5) and optimization area ($N_{A}$ $\sim$ 40-100) to suppress speckle noise, balanced against other settings to reduce self-subtraction: local masking and/or a higher $SVD_{\rm lim}$ cutoff (e.g. 10$^{-4}$ instead of 10$^{-6}$).}.

\subsection{Detections}
Figure \ref{fig:images} shows the detection of a faint point source, HIP 54515 b (denoted by an arrow), at the 10--11 o'clock position about 1--2 $\lambda$/D away from the coronagraph mask edge in each epoch.  Following previous work to calculate the companion signal-to-noise ratio (SNR) with a finite-element correction \citep{Currie2011,Mawet2014}, the HIP 54515 b SNR ranges between 5.4 and 16.2 in the wavelength-collapsed CHARIS images\footnote{We calculated the SNR following standard methods.  We replaced each pixel by its sum within a FWHM-sized aperture and then computed the robust standard deviation of (convolved) pixels as a function of radial separation.   We applied a finite-element correction to the noise estimate following \citet{Mawet2014}: this correction is substantial at HIP 54515 b's separation ($\approx$1.3--2, depending on channel and epoch of observation).  Thus, the SNR quoted throughout this work is equal to that valid for gaussian noise with an infinite number of noise samples: e.g. SNR = 5 yields a false-alarm probability of $\sim$2.86$\times$10$^{-7}$.}
In the highest-quality data sets (December 2022, February 2024, May 2025), HIP 54515 b is easily visible in each CHARIS channel.

\begin{figure*}[ht!]
   \includegraphics[width=0.33\textwidth,clip]{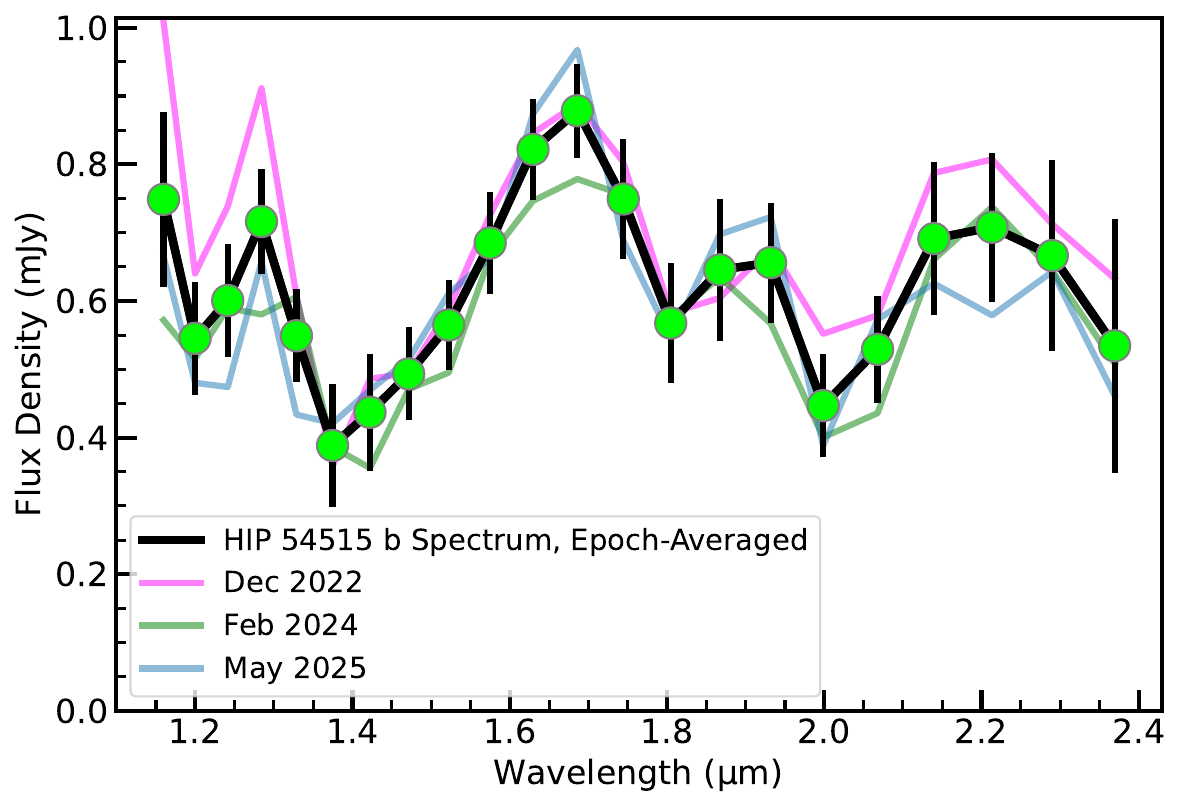}
   \includegraphics[width=0.33\textwidth,clip]{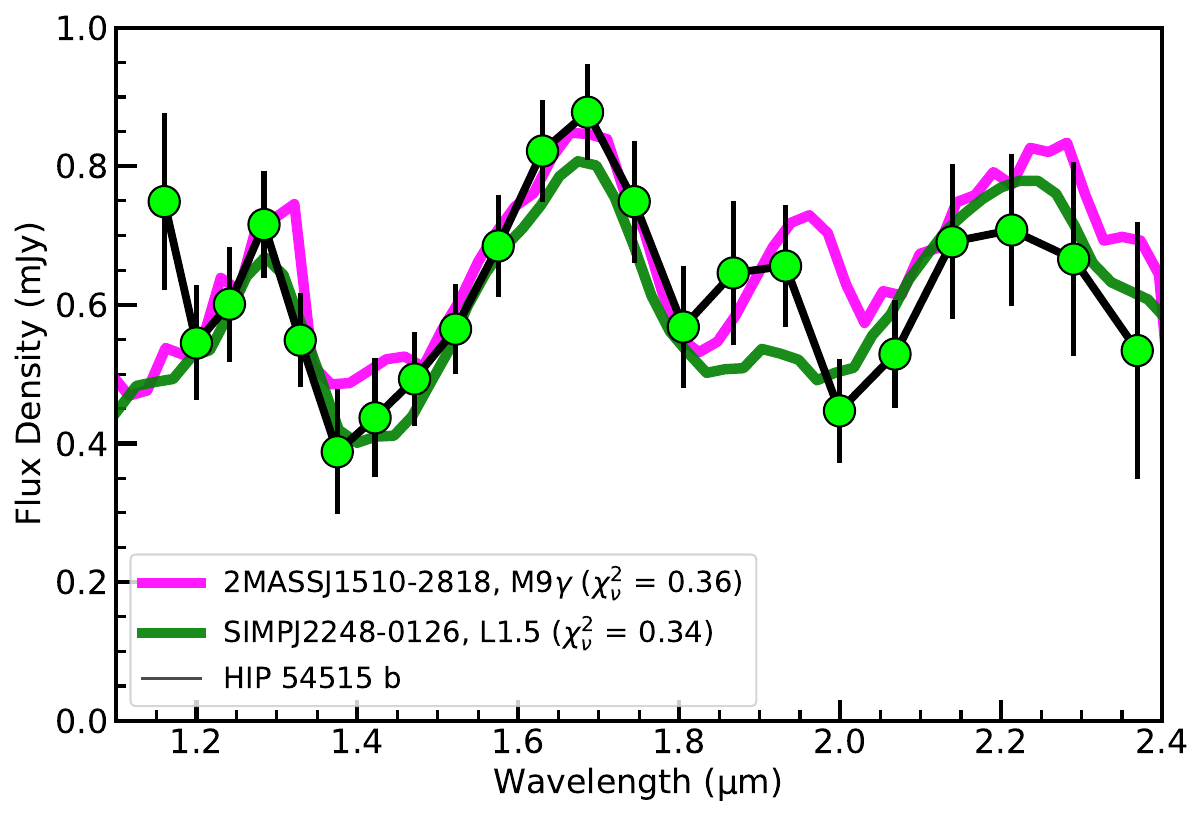}
   \includegraphics[width=0.33\textwidth,clip]{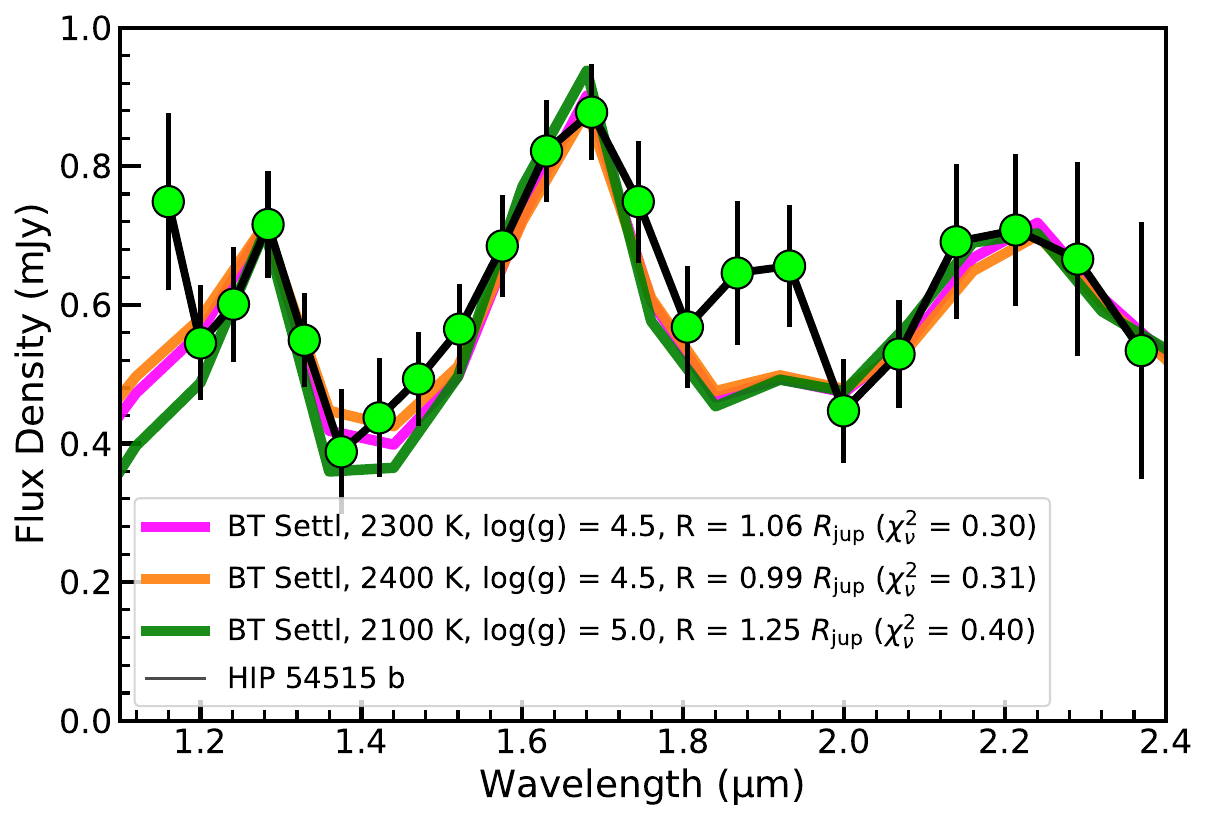}

  \caption{
Epoch-averaged HIP 54515 b spectrum compared to spectra extracted from individual epochs (left) and to selected best-fitting substellar objects in the Montreal Spectral Library (middle) and BT-Settl atmosphere models (right).
  }
  \label{fig:spectrum}
\end{figure*}

\section{Spectroscopic Analysis}
\label{sec:spec_analysis}
To extract a calibrated HIP 54515 b spectrum and astrometric measurements, we first forward-modeled the companion signal at its location to assess spectrophotometric and astrometric biasing \citep{Pueyo2016,Currie2018}.  The channel-averaged throughput for each epoch ranged between $\approx$0.2 and $\approx$0.5.  Astrometric biasing was typically $\sim$0.05--0.1 CHARIS pixels along both coordinates on the detector.  From the fitted companion centroid position, we extracted its spectrum, correcting for signal loss in each channel.   Spectrophotometric errors consider the intrinsic SNR of the detection and errors in the absolute spectrophotometric calibration \citep[see ][]{Currie2020a}. 

Following \citet{GrecoBrandt2016}, we compute the spectral covariance at HIP 54515 b's position.   In the December 2022 epoch, the noise is even split between spatially/spectrally-correlated noise and uncorrelated noise (i.e. $A_{\rm \rho}$ $\sim$ 0.45, $A_{\rm \lambda}$ $\sim$ 0.1, and $A_{\rm \delta}$ $\sim$ 0.45).  In the February 2024 and May 2025 epochs, the noise is slightly more uncorrelated ($A_{\rm \delta}$ $\sim$ 0.57 and $A_{\rm \delta}$ $\sim$ 0.72, respectively).

Figure \ref{fig:spectrum} (left panel) shows HIP 54515 b's spectrum extracted from our three highest-quality datasets and the average value for each epoch (Table \ref{tab:spec}).   HIP 54515 b exhibits a ``sawtooth" spectrum characteristic of substellar objects.  Measurements from different epochs show the best agreement at $H$ band.  The photometry derived from the epoch-averaged spectrum is $J$, $H$, $K$ = 15.99 $\pm$ 0.14, 15.38 $\pm$ 0.11, 15.08 $\pm$ 0.17 in the Mauna Kea Observatories filter set: values from the three individual epochs are consistent to within $\sim$1-$\sigma$.

\begin{deluxetable}{llll}
     \tablewidth{0pt}
    \tablecaption{HIP 54515 b Spectrum (Average)\label{tab:spec}}
    \tablehead{\colhead{Wavelength ($\mu$m)} & \colhead{$F_{\rm \nu}$ (mJy)} &  \colhead{$\sigma$~$F_{\rm \nu}$ (mJy)} & \colhead{SNR}}
    \startdata
1.160 & 0.749 & 0.128 & 8.2 \\
1.200 & 0.545 & 0.083 & 7.2 \\
1.241 & 0.601 & 0.083 & 7.4 \\
1.284 & 0.716 & 0.077 & 9.6 \\
1.329 & 0.549 & 0.068 & 8.8 \\
1.375 & 0.388 & 0.090 & 5.0 \\
1.422 & 0.437 & 0.086 & 5.4 \\
1.471 & 0.493 & 0.068 & 7.8 \\
1.522 & 0.565 & 0.065 & 9.6 \\
1.575 & 0.685 & 0.074 & 10.0 \\
1.630 & 0.822 & 0.074 & 11.8 \\
1.686 & 0.878 & 0.069 & 14.3 \\
1.744 & 0.749 & 0.088 & 9.4 \\
1.805 & 0.568 & 0.088 & 8.7 \\
1.867 & 0.646 & 0.104 & 7.8 \\
1.932 & 0.656 & 0.088 & 7.9 \\
1.999 & 0.447 & 0.075 & 6.6 \\
2.068 & 0.529 & 0.078 & 7.0 \\
2.139 & 0.691 & 0.112 & 6.5 \\
2.213 & 0.708 & 0.109 & 6.8 \\
2.290 & 0.666 & 0.140 & 5.1 \\
2.369 & 0.534 & 0.186 & 3.5 \\
\enddata
\end{deluxetable}

Comparing HIP 54515 b's average spectrum to substellar objects in the Montreal Spectral Library \citep{Gagne2015} reveals that the companion is likely near the M/L transition, Figure \ref{fig:spectrum} (middle panel).   Comparisons with many M5--L5 objects formally yield $\chi_{\nu}^{2}$ $\lesssim$1 due to HIP 54515 b's relatively large spectroscopic uncertainties.  However, best-fitting objects are predominantly low-gravity M8-M9 dwarfs or field L1-L2 dwarfs.  From \citep{Pecaut2013}, this range of spectral types favors temperatures of $\approx$ 2000-2600 $K$.  

We use the solar metallicity BT-Settl model grid to further explore HIP 54515 b's atmosphere \citep{Allard2012}, covering temperatures and gravities of 400-4000 K and log(g) = 3--5.5, respectively.    Our fitting procedure follows that outlined in \citet{ElMorsy2024b}, considering the CHARIS spectral covariance\footnote{We fit the epoch-averaged spectrum and use the epoch-averaged covariance, but we find consistent results using covariances from individual epochs.}.  

The best-fitting models cover 2200-2400 $K$, a wide range of gravities, and radii range from $\sim$0.91 to $\sim$1.27 $R_{\rm Jup}$ (Figure \ref{fig:spectrum}, right panel).  Considering models fulfilling $\chi^{2}$ $\le$ $\chi^{2}_{\rm min}$ + 6.18 (i.e. a joint 2-$\sigma$ confidence interval in temperature and gravity), HIP 54515 b's best-estimated luminosity, temperature, gravity, and radius are then log($L/L_{\odot}$) $\approx$ -3.52 $\pm$ 0.03, $T_{\rm eff}$ $\approx$ 2348 $\pm$ 218 $K$, log($g$) $\approx$ 4.46 $\pm$ 0.81, and $R$ $\approx$ 1.08 $\pm$ 0.23 $R_{\rm Jup}$. We obtain consistent results with the Sonara Diamondback and Exo-REM model grids \citep{Charnay2018,Morley2024}.

\section{Astrometric Analysis}
\label{sec:astrometric_analysis}

\subsection{Calibrated Astrometric Measurements}
Astrometric uncertainties consider the centroiding uncertainty of the companion, uncertainty in the star's position, and uncertainties in the north position angle and pixel scale.  As described in Appendix \ref{astrocal}, we recalibrated CHARIS's astrometric solution using CHARIS astrometry for M5 and $\theta$ Ori 1 B, matched to astrometry derived from archival Keck/NIRC2 data.  We adopt a pixel scale of 16.10 mas pixel$^{-1}$ $\pm$ 0.04 mas pixel$^{-1}$ and a north position angle offset of -2.03$^{o}$ $\pm$ 0.27$^{o}$, which is formally consistent with prior results based on binaries or a combination of binaries cross-calibrated with Keck/NIRC2 and early M5 data pinned to Hubble Space Telescope data from \citet{Currie2018} and \citet{Chen2023}.

To assess astrometric biasing due to PSF subtraction and compute astrometric uncertainties, we first compared the imputed position of the HIP 54515 b forward-modeled PSF with the recovered position of the forward-modeled PSF after PSF subtraction to estimate the astrometric biasing.  We then computed the scatter in recovered positions of the planet forward-model -- injected into the final PSF-subtracted cube and scaled to null the real signal -- at the same angular separation as HIP 54515 b but different position angles to estimate the centroiding uncertainty due to residual speckle noise.

Table \ref{astrom} lists HIP 54515 b's astrometry for each epoch.  We adopt a stellar centroiding uncertainty of 0.25 pixels, a pixel scale uncertainty of 0.05 mas, and a north position angle uncertainty of 0.27$^{o}$ \citep{Currie2018}.  We converted the HIP 54515 b astrometry corrected for biasing into polar coordinates, considering all astrometric uncertainties in quadrature.   Typical astrometric errors are $\sim$4 mas in angular separation and $\sim$ 1$^{o}$ in position angle, dominated by the intrinsic detection SNRs.  The first epoch has the largest astrometric uncertainty because it has a greater scatter in the recovered planet forward-model positions, likely due to greater correlated noise.   A background star originally located at HIP 54515 b's position in the 2022 December data would appear at $\rho$ $\sim$ 0\farcs{}317 and a position angle of 76.83$^{o}$ in our 2025 May data due to the star's proper motion (PM$_{\alpha}$, PM$_{\delta}$ $\sim$ [-65.160, -5.525] mas yr$^{-1}$).  Such an object is discrepant with HIP 54515 b's 2025 May astrometry by over 35-$\sigma$.

\begin{deluxetable}{ccc}[!h]
 \label{astrom}
     \tablecaption{HIP 54515 b Astrometry}
      \tablehead{\colhead{UT Date} & \colhead{$\rho$ ($\arcsec{}$)} & \colhead{PA (deg)}}
     \startdata
    20221230 &  0.145 $\pm$ 0.007 & 64.813 $\pm$ 1.171\\
    20240220  & 0.174 $\pm$ 0.004 & 56.204 $\pm$ 0.984\\
    20240427  & 0.183 $\pm$ 0.005 & 53.456 $\pm$ 0.947\\
    20250315 & 0.193 $\pm$ 0.004 & 49.477 $\pm$ 0.901\\
    20250512 & 0.191 $\pm$ 0.004 & 47.329 $\pm$ 0.907
    \enddata
\end{deluxetable}

   \begin{deluxetable*}{lccr}[]
\tablecaption{MCMC Orbit Fitting Priors and Results}
\tablewidth{0pt}
  \tablehead{
  \colhead{Parameter} & \colhead{16/50/84\% quantiles} & Prior}
  \startdata
\multicolumn{3}{c}{Fitted Parameters (1,2)$^{a}$} \\ \hline
RV jitter (m/s)  &      ${0.095}_{-0.095}^{+50}$, ${0.12}_{-0.12}^{+52}$ & log-uniform \\
$M_{\rm pri}$ ($M_\odot$)   &      ${1.90}_{-0.20}^{+0.19}$, ${1.92}_{-0.20}^{+0.19}$  & Gaussian, $1.8 \pm 0.2$ \\
$M_{\rm sec}$ ($M_{\rm Jup}$)  &      ${17.7}_{-4.9}^{+7.6}$, ${19.2}_{-5.2}^{+9.4}$  & 1) $1/M_{\rm sec}$$^{a}$ (log-uniform), 2) flat$^{a}$ \\
Semimajor axis $a$ (au)    &     ${24.8}_{-4.7}^{+7.2}$, ${25.0}_{-4.8}^{+7.7}$ & $1/a$  (log-uniform) \\
$\sqrt{e} \sin \omega$    &    ${-0.31}_{-0.32}^{+0.87}$,${-0.37}_{-0.26}^{+0.91}$ & uniform \\
$\sqrt{e} \cos \omega$  &        ${-0.09}_{-0.39}^{+0.43}$, ${-0.05}_{-0.38}^{+0.38}$  & uniform  \\
Inclination ($^\circ$)    &      ${129.6}_{-5.9}^{+6.7}$, $129.1^{+7.1}_{-6.1}$   &  $\sin i$ (geometric) \\
PA of the ascending node $\Omega$ ($^\circ$)  &      ${33}_{-20}^{+173}$, ${31}_{-17}^{+174}$ & uniform \\
Mean longitude at 2010.0 ($^\circ$) &        ${197}_{-158}^{+43}$, ${206}_{-163}^{+41}$ & uniform \\
Parallax (mas)   &      ${12.0389}_{-0.0062}^{+0.0060}$, ${12.0389}_{-0.0061}^{+0.0060}$ & Gaussian, 0 \\
\hline
\multicolumn{3}{c}{Derived Parameters(1,2)$^{a}$} \\ \hline
Period (yrs)     &    ${89}_{-24}^{+41}$, ${91}_{-26}^{+45}$ \\
Argument of periastron $\omega$ ($^\circ$)   &      ${214}_{-150}^{+56}$, ${228}_{-155}^{+52}$ \\
Eccentricity $e$    &   ${0.42}_{-0.14}^{+0.13}$, ${0.41}_{-0.15}^{+0.13}$   \\
Semimajor axis (mas)  &  ${297}_{-55}^{+85}$, ${303}_{-59}^{+92}$ \\
Periastron time $T_0$ (JD)    &    ${2470981}_{-10740}^{+13757}$, ${2471350}_{-10967}^{+14701}$\\
Mass ratio   &      ${0.0090}_{-0.0024}^{+0.0036}$, ${0.0097}_{-0.0026}^{+0.0047}$
\enddata
\tablenotetext{a}{Quantiles listed are for the simulations assuming (1) a log-uniform prior and (2) a flat prior. }
\label{tab:mcmc_result}
\end{deluxetable*}


\begin{figure*}
\centering
   \includegraphics[width=\textwidth]{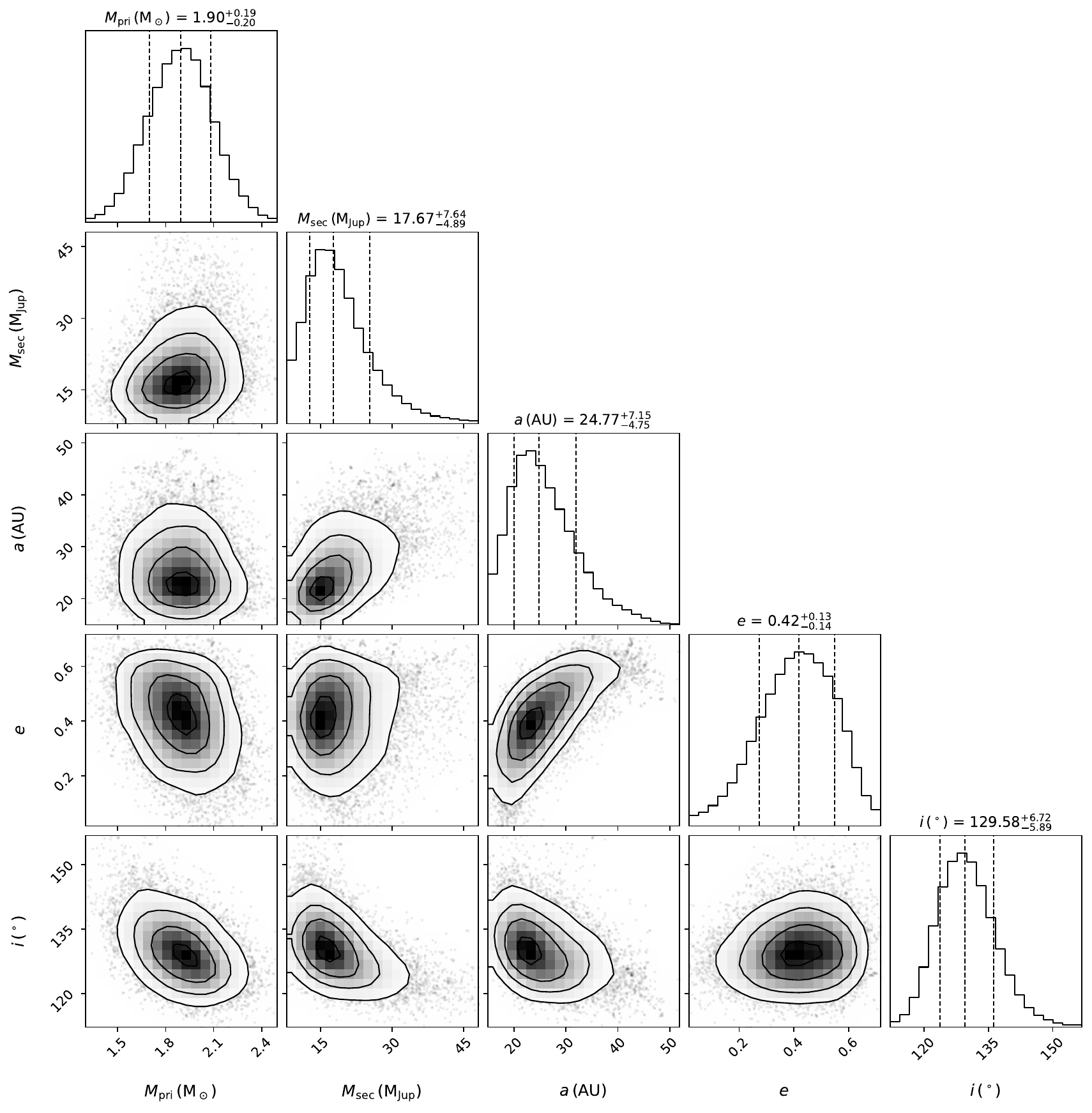}
   \caption{Corner plot of the downsampled posterior distributions for HIP~54515~b assuming a log-uniform prior on the companion mass. The plot shows marginalized 1D and 2D posteriors for the primary mass ($M_{\rm pri}$), companion mass ($M_{\rm sec}$), semi-major axis ($a$), eccentricity ($e$), and inclination ($i$). The contours represent the 68\%, 95\%, and 99.7\% credible regions.  Note that the downsampling uniformly selects 10,000 random samples from the flattened, post-burn-in posterior to produce corner plots without altering the underlying distribution. }
   \label{fig:corner_lognormal} 
\end{figure*}

\begin{figure}[ht]
\centering
\includegraphics[width=0.39\textwidth]{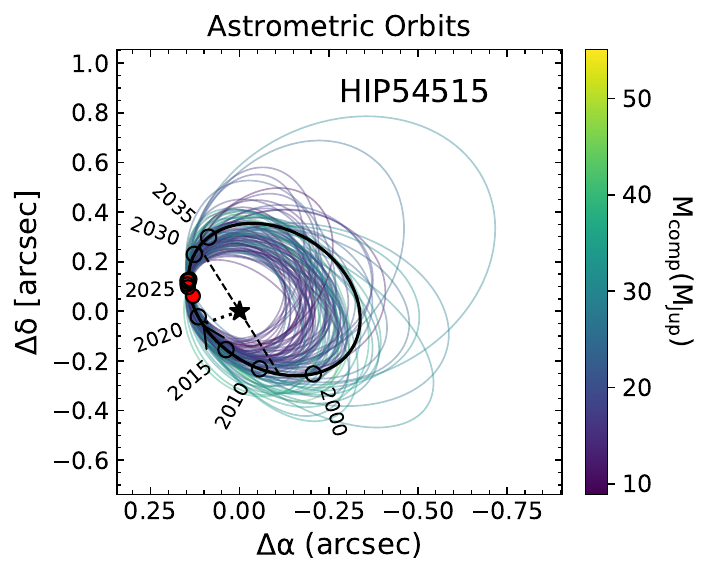}\\ \hspace*{+2.5mm}
\includegraphics[width=0.365\textwidth]{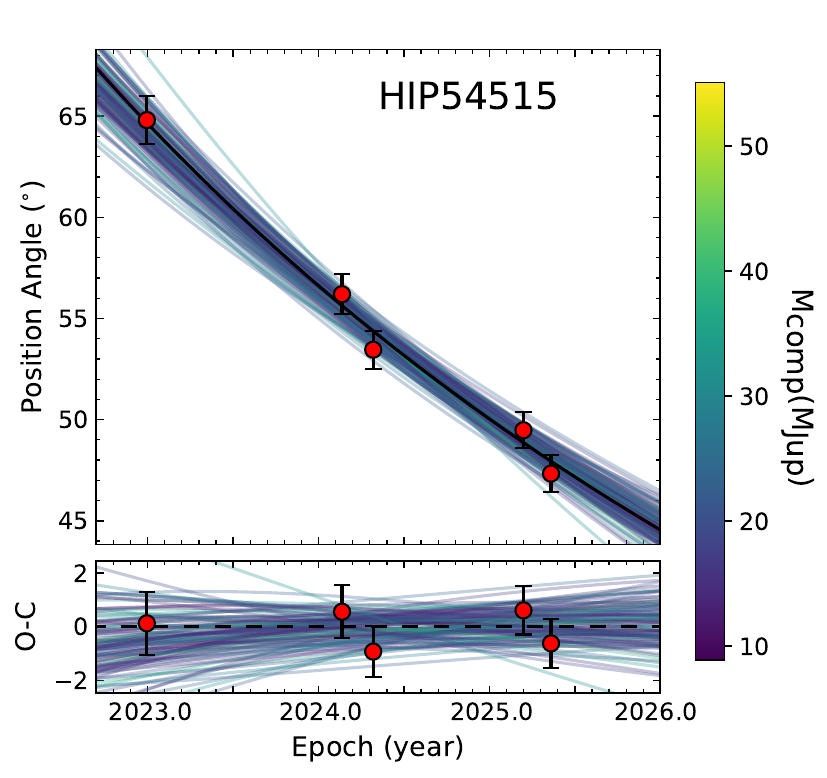}\\
\includegraphics[width=0.387\textwidth]{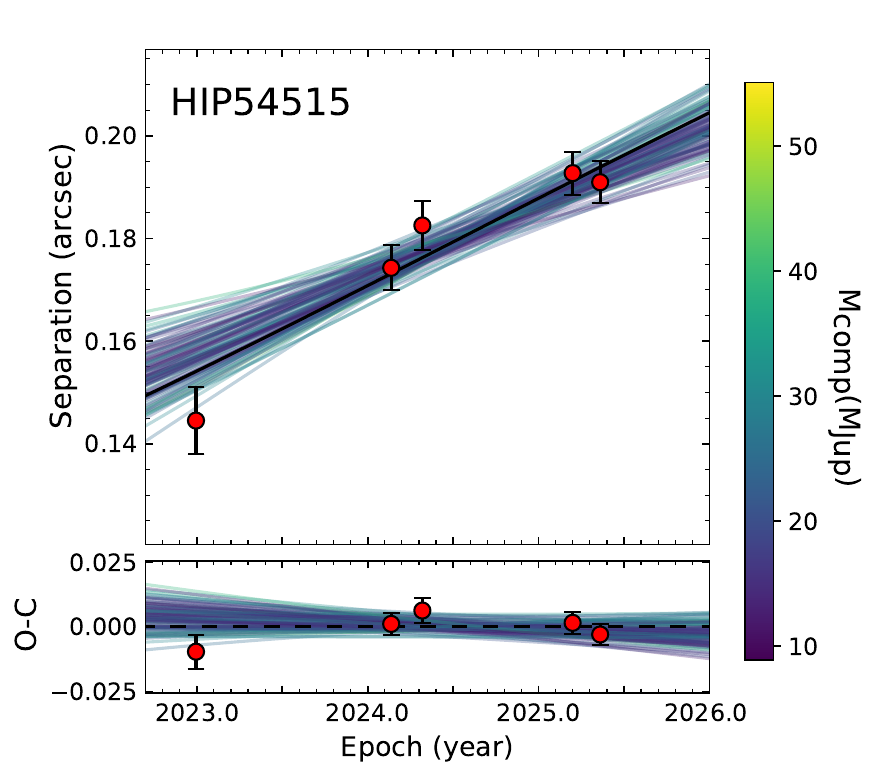}\\
\caption{Orbital fit for HIP~54515~b assuming a log-uniform prior on companion mass using \texttt{orvara}.
(Left) Ensemble of orbital solutions with the maximum likelihood orbit shown in black and 70 randomly drawn posterior samples overplotted and color-coded by companion mass. Filled red circles indicate our new CHARIS astrometric data, while open circles mark the predicted locations at each epoch. (Center) Position angle as a function of time. (Right) Angular separation versus epoch. All panels reflect the range of plausible orbits consistent with the data and inferred parameter uncertainties.}
\label{fig:hip54515_orbits_lognormal}
\end{figure}

\subsection{Constraining HIP 54515 b's Dynamical Mass and Orbit}
We use the Bayesian orbital retrieval software \texttt{orvara} to derive the dynamical mass and orbit of HIP~54515~b by combining our CHARIS relative astrometry with Hipparcos-Gaia (eDR3) proper motion accelerations \citep{Brandt2021}.   We run two simulations: one adopting a standard log-uniform prior (i.e. 1/$M_{\rm sec}$) for HIP 54515 b's mass and another adopting a flat mass prior\footnote{As noted in \citet{Bovie2025}, the different priors are connected to different assumptions about the underlying frequency.}
of companions as a function of mass over the range relevant for HIP 54515 b.   The mass function decreases with increasing mass below 25 $M_{\rm Jup}$, slightly increases from $\approx$50 $M_{\rm Jup}$ to the hydrogen-burning limit, and turns over somewhere between 25 $M_{\rm Jup}$ and 40 $M_{\rm Jup}$ \citep[e.g.][]{Sahlmann2011,Kiefer2019,Currie2023b,Stevenson2023}. While a log-uniform prior may more accurately represent the underlying distribution for plausible masses for HIP 54515 b (see the results below), a flat prior is a more conservative assumption.  For other parameters, we adopt standard priors, following Table 4 in \citet{Bovie2025}. 
Both simulations run \texttt{orvara} with 20 temperatures, 100 walkers for each temperature, and 200,000 steps per walker. We thinned the chain by a factor of 50 and treated the first 100 steps of the thinned chain as burn-in. We verify that both chains used for inference have converged after burn-in based on the autocorrelation criterion. 

The posterior estimates from our orbit modeling using the log-uniform prior and flat prior are summarized in Table~\ref{tab:mcmc_result}. The log-uniform prior simulation yields a companion mass of ${17.7}_{-4.9}^{+7.6}\,M_{\rm Jup}$, a primary mass of ${1.90}_{-0.20}^{+0.19}\,M_{\rm \odot}$, and thus a mass ratio of ${0.0090}_{-0.0024}^{+0.0036}$.  HIP 54515 b's semimajor axis $a$ is $24.9^{+5.5}_{-4.2}$ au, with an eccentricity of ${0.42}_{-0.14}^{+0.13}$ and inclination of ${129.6^{+6.7}_{-5.9}}^{\circ}$. In comparison, the flat prior results in a marginally higher but consistent companion mass of ${19.2}_{-5.2}^{+9.4}\,M_{\rm Jup}$, a similar semimajor axis of ${25.0}_{-4.8}^{+7.7}$ au, and nearly identical eccentricity of ${0.41}_{-0.15}^{+0.13}$. The inclination remains consistent at ${129.1^{+7.1}_{-6.1}}^{\circ}$, with a marginally higher mass ratio of ${0.0097}_{-0.0026}^{+0.0047}$. The two posteriors are consistent within $0.2\sigma$ for all parameters\footnote{Other slight modifications to our approach yield consistent results.  For example, setting ``epoch astrometry = True" to take advantage of intermediate astrometric data from Hipparcos and the Gaia Observation Forecast Tool \citep{Brandt2021e} results in companion mass posteriors of ${17.2}_{-4.7}^{+8.0}$ $M_{\rm Jup}$and ${19.7}_{-5.6}^{+10}$ $M_{\rm Jup}$ for a log-uniform and flat prior, respectively. }

Figure~\ref{fig:corner_lognormal} shows the posterior distributions under the log-uniform prior on companion mass, while Figure~\ref{fig:hip54515_orbits_lognormal} displays representative orbits sampled from the posterior. 
For simplicity, we include only the corner plot and orbital sampling for the log-uniform case.
Most parameters in Figure~\ref{fig:corner_lognormal} exhibit well-constrained, single-peaked posterior distributions. The posterior for $M_{\rm sec}$ is slightly skewed, with a tail toward higher masses, consistent with the log-uniform prior and limited orbital coverage. Several parameter pairs ($M_{\rm sec}$ and the semi-major axis $a$, $a$ and $e$) exhibit a mild covariance: lower-mass companion solutions slightly favor tighter, lower-eccentricity orbits. For the flat prior on companion mass, the corner plot (not shown) is nearly identical to the log-uniform case, except for marginally higher masses (8.6\% increase in the median value).

The companion mass posterior distributions from both simulations favor values centered on 17.7--19.2 $M_{\rm Jup}$ with 68\% confidence intervals strongly disfavoring  masses above $\sim$ 25 $M_{\rm Jup}$ and excluding those above 30 $M_{\rm Jup}$.  At these masses, prior studies find that the companion mass function is not flat but is decreasing with mass \citep[][see also Section 5.3 in this work]{Feng2022,Xiao2023,Currie2023a}.  The log-uniform prior then better represents the intrinsic distribution of substellar companion masses from which HIP 54515 b is ostensibly drawn.  As a consequence, we favor the log-uniform simulations dynamical parameters.

\subsection{HIP 54515 b Analyzed within the Context of Other Substellar Companions: A Directly Imaged Superjovian Planet}

The criteria best supported in the peer reviewed literature to determine what kind of object HIP 54515 b might be -- planet or brown dwarf -- has undergone significant revision. 
Some studies continue to distinguish between planets and brown dwarfs based on an object's mass relative to the deuterium-burning limit (DBL).  
However, the peer-reviewed literature over the past 15 yr analyzing substellar companion demographics does \textit{not} support the DBL criterion for distinguishing between planets and brown dwarfs \citep{Sahlmann2011,Kratter2010,Currie2011,Spiegel2011,Kiefer2019,Feng2022,Xiao2023,Stevenson2023,Currie2023a,Currie2023b,Meyer2025}   \footnote{
The DBL is ostensibly analogous to the hydrogen-burning limit at $\approx$75--80 $M_{\rm Jup}$, which differentiates between stars and brown dwarfs.  But this analogy fails.  Unlike hydrogen burning, deuterium burning's impact on the long-term luminosity evolution of a substellar object is negligible \citep[][]{Burrows2001}: there is no ``deuterium burning main sequence" or post main sequence, let alone one present in 20 $M_{\rm Jup}$ objects that is absent in 10 $M_{\rm Jup}$ objects.  

Furthermore, as noted by \citet{Spiegel2011}, objects well below the classical threshold of $\sim$13 $M_{\rm Jup}$ do burn some deuterium, and young ($t$ $\lesssim$ 100 Myr) objects up to $\approx$ 20 $M_{\rm Jup}$ are predicted to have deuterium abundances similar to Gyr-old 13 $M_{\rm Jup}$ objects.  The DBL is also claimed to be an observationally based criterion.  However, the exact mass limit at which substellar objects burn 50\% of their deuterium depends on the object's helium abundance, which is not directly observable nor can a reliable helium abundance be inferred from common spectral features in substellar objects.  See \citet{Luhman2008} and \citet{Currie2023b,Currie2023a} for further discussion.}.

Instead, the demographics of other substellar companions provide a context for assessing whether HIP 54515 b is best understood as part of a population of (super-)Jovian planets formed in a protoplanetary disk 
or brown dwarfs formed by molecular cloud fragmentation (i.e. like stars) \citep[][]{Luhman2008}.
Recent work based on the companion mass function inferred from the California Legacy Survey \citep{Rosenthal2021} and NASA Exoplanet Archive data favors a planet companion mass function that extends well past the classical deuterium-burning limit to $\sim$25--39 $M_{\rm Jup}$, broadly consistent with prior smaller studies \citep[e.g.][]{Sahlmann2011}, but also suggests that the companion-to-primary mass ratio ($q$) may be a more fundamental property distinguishing planets from brown dwarfs  \citep{Currie2023a}.  \citet{Currie2023b} combine multiple metrics to identify provisional, demographics-driven criteria for a planet of $M$ $\le$ 25 $M_{\rm Jup}$, log($q$) $\le$ -1.6, and $a$ $\le$ 300 au.  By these criteria, HIP 54515 b lies within the phase space of objects properly considered to be superjovian planets.

\begin{figure*}
\centering
\includegraphics[width=0.8\textwidth,clip]{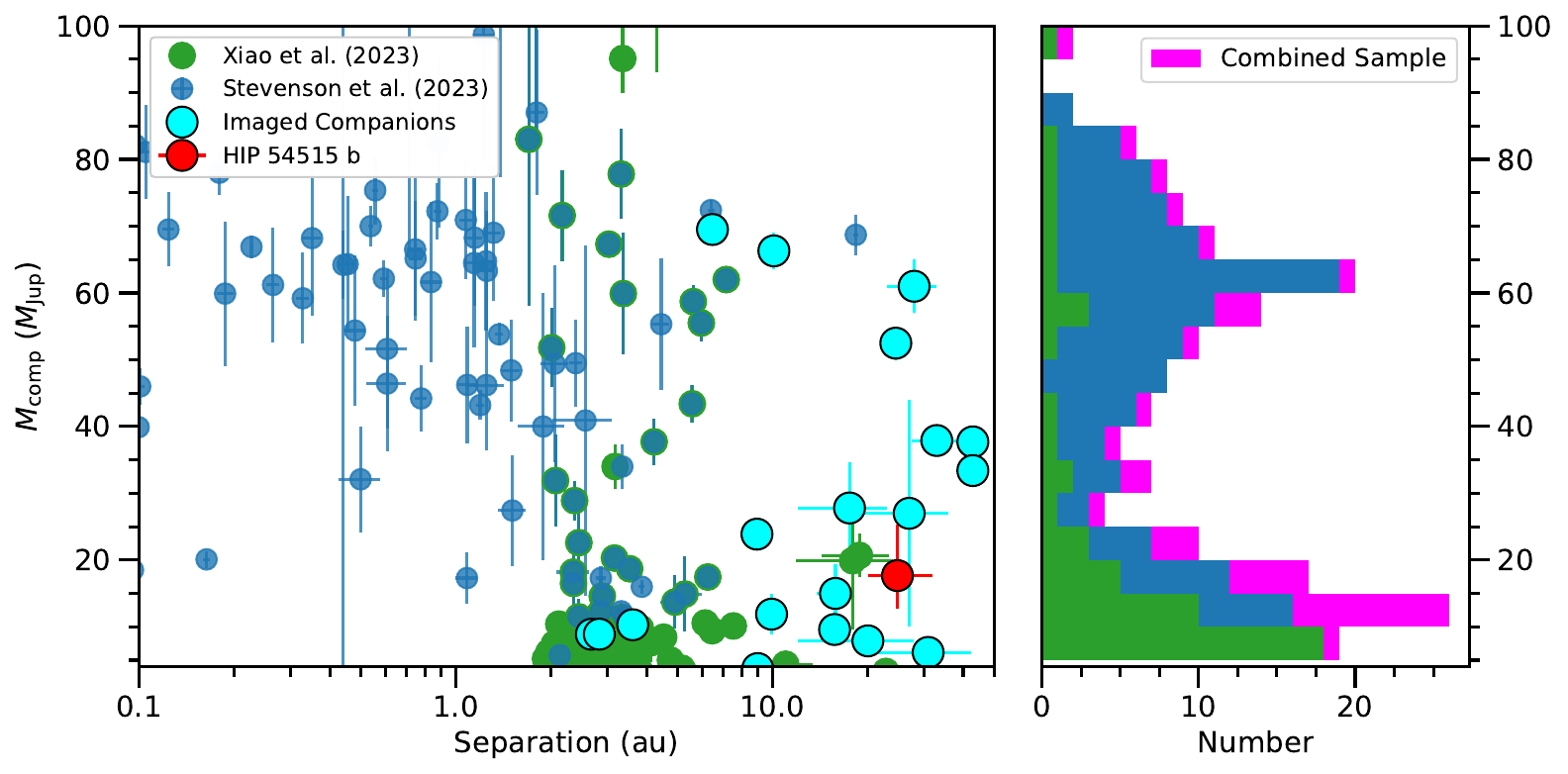}
\includegraphics[width=0.8\textwidth,clip]{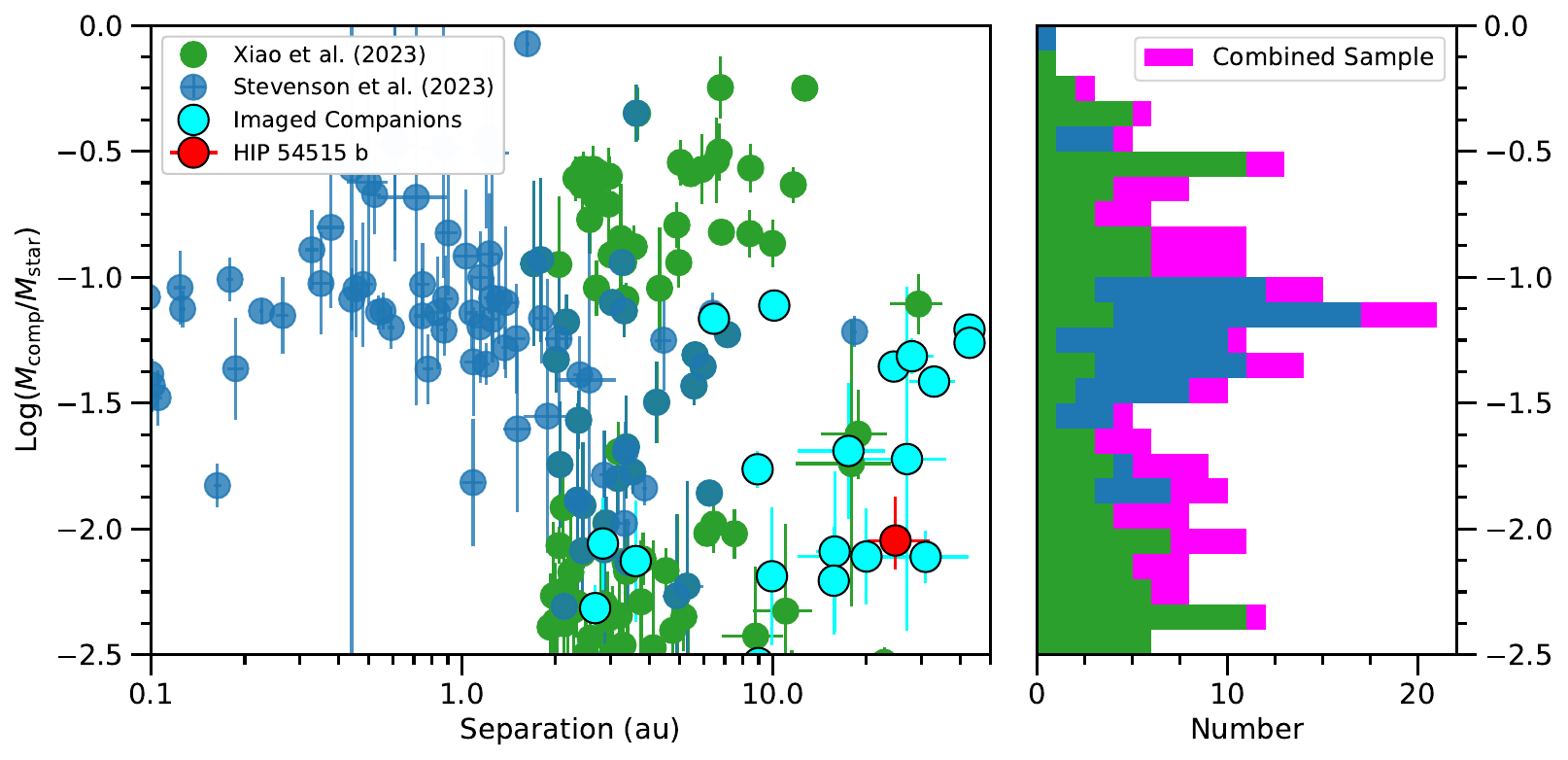}
\vspace{-0.125in}
   \caption{(Top) Mass versus semimajor axis and (bottom) mass ratio versus semimajor axis distributions from recent studies yielding dynamical masses for jovian planets at 0.1--50 au from \citet{Stevenson2023} and \citet{Xiao2023}.  To these samples, we add directly imaged substellar companions with dynamical masses: HIP 99770 b \citep{Bovie2025}, $\beta$ Pic bc \citep{Lacour2021}, AF Lep b \citep{Balmer2025}, HD 206893 B and c \citep{Sappey2025}, $\epsilon$ Ind b \citep{Matthews_2024}, 14 Her bc \citep{Bardalez2025}, PZ Tel B \citep{Franson2023b}, HD 984 B \citep{Franson2022}, GJ 758 B \citep{Brandt2019}, HD 4747 B \citep {Brandt2019}, HD 72946 B \citep{Balmer2023}, GJ 229 Ba and GJ 229 Bb \citep{Thompson2025}, HIP 21152 B \citep{Kuzuhara2022}, and HR 8799 e \citep{Brandt2021d}.  The histogram plots (right-hand panels) show the mass and mass ratio distributions for the individual samples and the combined sample. }
   \label{fig:massauplot} 
\end{figure*}

Here, we use the recent surveys from \citet{Xiao2023} and \citet{Stevenson2023} for an additional demographics context from which to interpret HIP 54515 b.  The samples in these studies are smaller than those used in \citet{Currie2023b} and cover a narrower range in semimajor axis.  \citet{Stevenson2023} may undersample planets less massive than $\approx$3--5 $M_{\rm Jup}$ due to astrometric sensitivity limits, while \citet{Xiao2023} heavily relies on the NASA Exoplanet Archive and exoplanet.eu, which focus on objects below 30 $M_{\rm Jup}$ and 50 $M_{\rm Jup}$, and thus may undersample the population of massive brown dwarfs near the hydrogen-burning limit as a result.   However, these studies' key advantage is that they yield dynamical masses, not lower limits (m~sin($i$)), using analysis tools similar/identical to those employed here.   

Figure \ref{fig:massauplot} (top panel) displaying mass vs. semimajor axis clearly reveals two populations drawn from these samples, pointing towards a brown dwarf population extending to the hydrogen burning limit and a planet population extending down to Jupiter masses.  For both samples, the planet population extends to $\sim$ 25 $M_{\rm Jup}$ at the high-mass end.  For the \citet{Stevenson2023} sample and combined sample, the low-mass end of the brown dwarf population extends to $\sim$ 30--40 $M_{\rm Jup}$: the samples include few 25--40 $M_{\rm Jup}$ objects.   The distributions of companion-to-primary-mass ratios likewise reveal a brown dwarf population with log($M_{\rm comp}$/$M_{\star}$) $>$ -1.5 and planet population with log($M_{\rm comp}$/$M_{\star}$) $<$ -1.6--1.7, separated at log($M_{\rm comp}$/$M_{\star}$) $\sim$ -1.6 ($M_{\rm comp}$/$M_{\star}$ $\sim$ 0.025) (bottom panel).   

While a clearer understanding of the turnover point in the companion mass and mass ratio functions from brown dwarfs to planets awaits more extensive future work, HIP 54515 b safely lies within the planet distribution for both parameters.   Other directly imaged planets -- e.g. $\beta$ Pic bc, HR 8799 e, HIP 99770 b, $\epsilon$ Ind b, and 14 Her c -- have comparable masses and/or mass ratios.  Thus, current evidence points to HIP 54515 b being a classified as a superjovian planet, not a brown dwarf.

\begin{figure*}
\centering
\includegraphics[width=0.38\textwidth,clip]{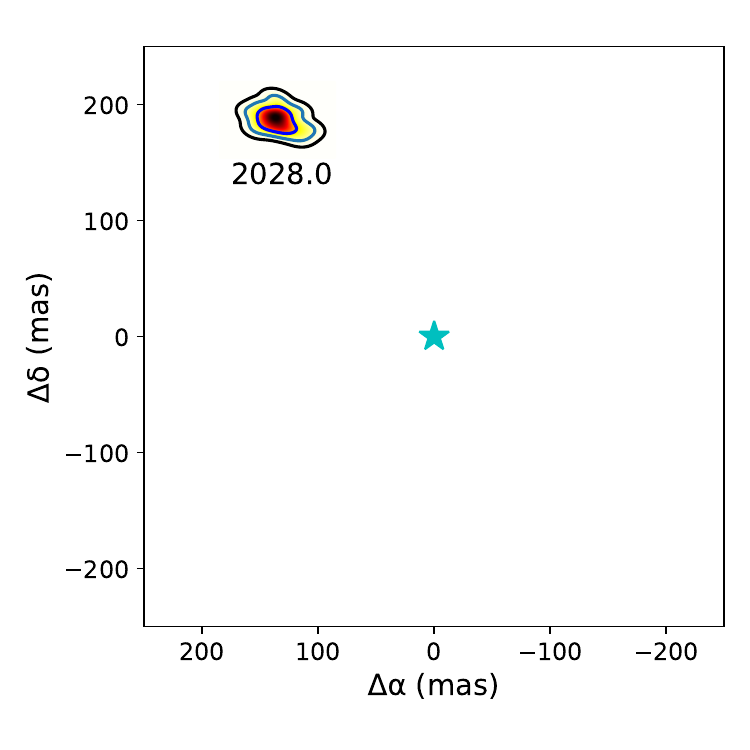}
\raisebox{3.5mm}{
\includegraphics[width=0.58\textwidth,clip]{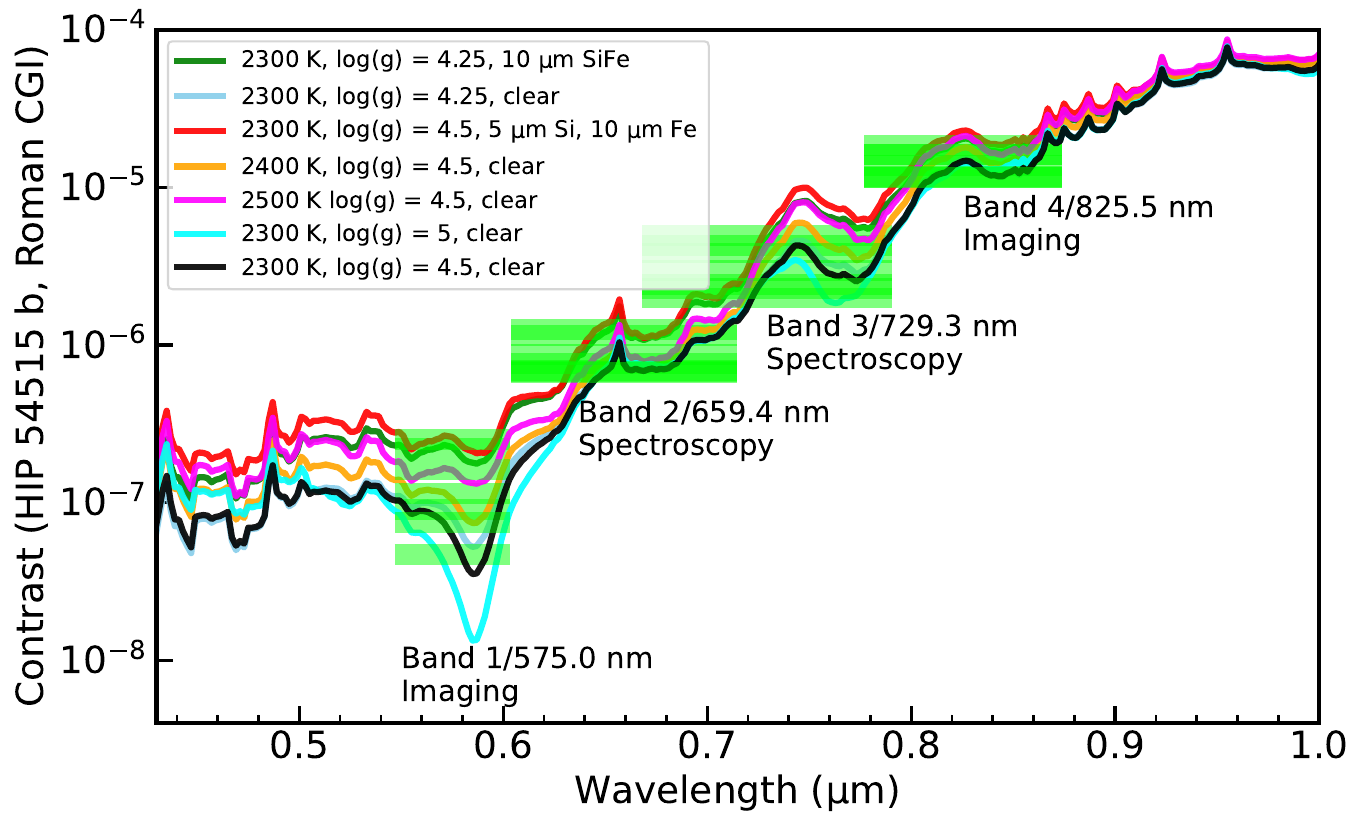}}
\vspace{-.25in}
   \caption{(Left) Predicted location (1, 2, 3 $\sigma$ contours) of HIP~54515~b relative to the host star (blue) from the best-fitting orvara orbits (assuming log-uniform priors on the companion mass) in January 2028. (Right) Predicted contrast in major CGI passbands for HIP 54515 b assuming a range of Lacy/Burrows cloudy and clear atmosphere models fitting the planet's CHARIS spectrum to within $\chi^{2}_{\rm \nu}$ $\le$ 1.   Models with clouds assume silicate and iron particles with modal sizes of 5 $\mu m$ or 10 $\mu m$.  The horizontal green bars show the predicted CGI contrasts for these models in the four CGI passbands.}
   \label{fig:cgi_prediction} 
\end{figure*}
\section{Discussion}
\label{sec:discussion}
\subsection{Summary of Results}
HIP~54515~b is the third superjovian planet jointly discovered using direct imaging and astrometry, following HIP~99770~b and then AF~Lep~b. It further showcases the power of pairing high-precision relative astrometry with HGCA proper motion accelerations to identify and dynamically characterize exoplanets, even at very small angular separations.   Moreover, unlike HIP~99770 and AF~Lep, the system had never been targeted for planet searches with any technique before.  Thus, precision astrometry from Gaia and Hipparcos may identify young, overlooked stars with detectable planets. 

\subsection{Further Characterization and Implications for Exoplanet Formation}
Our analysis estimates that HIP 54515 b's dynamical mass is ${17.7}_{-4.9}^{+7.6}\,M_{\rm Jup}$ and the system's age is $115^{+85}_{-91}$~Myr, making HIP 54515 b marginally more massive than HIP 99770 b but otherwise a younger analogue.  Like HIP 99770 b, HIP 54515 b also orbits a mid-A star.  HIP 54515 b's atmosphere appears to probe the M/L transition, in contrast to HIP 99770 b's location at the L/T transition.  
Follow-up spectroscopy with CHARIS at a slightly higher-resolution in individual $JHK$ passbands will better clarify HIP 54515 b's temperature, clouds, chemistry, and gravity as has been recently done for HIP~99770~b \citep[see][]{Bovie2025}.
Combined together and with improved stellar age estimates, the two planets may provide a valuable test case for calibrating how massive jovian planet atmospheres evolve during the early stages of cooling (e.g., \citealt{Brandt2021c}).

Considered as a population, imaged jovian exoplanets are thought to have characteristically lower eccentricities than massive brown dwarfs \citep{Bowler2020,Nagpal2023}.   HIP~54515~b adds to a growing sample of newly discovered exoplanets with moderate eccentricities ($e$ $\approx$ 0.3--0.5), including $\beta$ Pic c, HIP~99770~b, and $\epsilon$~Indi~Ab \citep{Nowak2020,Lacour2021, Currie2023b, Matthews_2024,Bovie2025,Winterhalder2025}.  Additionally, the 51 Eri b exoplanet and multiple other previously known companions (e.g. $\kappa$ And b) have comparable-to-higher eccentricities \citep{Dupuy2022,Morris2024}. 

The mechanisms responsible for elevated planet eccentricities may be challenging to pinpoint.
Higher eccentricities can be consistent with formation via core accretion followed by dynamical interactions between planets (e.g. planet-planet scattering) or planets and a distant star (the Kozai-Lidov mechanisms) \citep{Takeda2005}.  Such dynamical interactions between $\beta$ Pic b and c may explain the latter's moderate eccentricity.  On the other hand, our imaging data and prior work do not identify a second, more massive companion around HIP 54515, HIP 99770, 51 Eri, and $\epsilon$ Ind, although the latter has been speculated \citep{Matthews_2024}.   The 2000 au separation binary companion to 51 Eri (GJ 3305) is unlikely to be responsible for 51 Eri b's elevated eccentricity via the Kozai-Lidov mechanism \citep{Montet2015}. Similarly, the Kozai-Lidov timescale for the $\sim$11,700 au separation brown dwarf binary $\epsilon$ Ind Ba/Bb to perturb $\epsilon$ Ind b is $\gtrsim$ 10 billion years for eccentricities below $\approx$ 0.995. 

As one alternative among others, massive planets may gain elevated eccentricities during the dispersal of eccentric protoplanetary disks with inner cavities (cleared by the planet), as described by \citet{LiLai2023}.  Adopting Eq. 33 from \citeauthor{LiLai2023}, a planet with HIP 54515 b's mass, current orbital period, and current semimajor axis could have its eccentricity excited through secular resonances with the disk if the cavity radius is $\approx$65--80 au, comparable in size to large cavities resolved with ALMA \citep{Francisvandermarel2020}.   A larger sample of planets with well-constrained masses, orbits, and eccentricities can further explore formation processes.

\subsection{HIP 54515 b as a Roman Coronagraph Technology Demonstration Target }
HIP~54515~b could also be strategically important for the CGI technology demonstration phase \citep{Kasdin2020,Bailey2023}.
Roman CGI is designed to deliver high-contrast imaging in reflected light at visible wavelengths (e.g., 575 nm), with a nominal dark hole between 0$.^{\prime\prime}$15 and 0$.^{\prime\prime}$45 around bright stars ($V \lesssim 5$). The CGI's primary tech demo requirement (TTR5) is to achieve raw contrasts of better than $10^{-7}$ within its dark hole on bright stars.  Secondary goals include obtaining photometry, astrometry, and spectroscopy of faint companions. Using our best-fit orbit from dynamical modeling with $\tt orvara$, we predict the position of HIP~54515~b at the start of the Roman mission (e.g., 2028 January). As shown in Figure~\ref{fig:cgi_prediction} (left),  HIP~54515~b is expected to lie at a mean projected separation of $\sim$0$^{\prime\prime}.23 \pm 0.^{\prime\prime}008$, and a mean position angle of $324.4 \pm 2.1^{\circ}$, within the CGI dark hole at both 575~nm and 730~nm.

To estimate HIP 54515 b's contrast in CGI passbands, we use an updated version of the \citet{LacyBurrows2020} presented in \citet{Bovie2025} and \citet{ElMorsy2024b}.   We focus on temperatures between 2300 K and 2500 K with gravities between log(g) = 4.25 and 5.  For the host star, we adopt a Kurucz atmosphere model appropriate for an A5V dwarf. At 575 nm, we estimate an optical contrast of $\approx$7$\times10^{-8}$ to 2.5$\times$10$^{-7}$, depending on the assumed temperature, gravity, and cloud properties (Figure \ref{fig:cgi_prediction}, right panel).   HIP 54515 b's 575 nm contrast is thus comparable to the  tech demo requirement of 10$^{-7}$ and a factor of 20--100 higher than its predicted performance ($\sim$3$\times$10$^{-9}$)
\footnote{\url{https://roman.ipac.caltech.edu/page/param-db\#coronagraph_mode}.   
End-to-end thermal vacuum chamber tests have achieved raw 5-$\sigma$ contrasts $<$3$\times$10$^{-8}$: see \url{https://workshop.ipac.caltech.edu/romancgi24/talks/Day1_Poberezhskiy.pdf}}.  At 730 nm, where CGI will perform long-slit spectroscopy, HIP 54515 b's predicted contrast is $\sim$2--5 $\times$10$^{-6}$.  

With a host star brightness of V$\approx$6.8, HIP~54515~b lies near the faint limit where CGI is expected to maintain effective wavefront control and high-contrast imaging performance. The companion is fainter than typical CGI targets, and there are systems where a CGI detection of a companion fulfilling the main TTR5 requirement is less challenging \citep[e.g.][]{ElMorsy2025}\footnote{HIP 54515 observations with Roman CGI also require a suitable PSF reference star.  The Roman Community Participation Program's vetting of candidate reference stars is still in progress.  However, we have identified multiple candidates from their list are suitable.  For example, HIP 57632 has a $\Delta$ pitch angle less than 3$^{o}$ for multiple $\sim$70-day periods in 2027 where keepout maps do not prevent it or HIP 54515 from being observable: unpublished SCExAO/CHARIS data fail to identify a companion within the Roman CGI dark hole region that would prevent it from being a suitable PSF reference (T. Currie, unpublished).  These and similar topics are further explored in \citet{Currie2025}.}.  However, HIP 54515 b provides a valuable test case for CGI performance in the challenging regime of small inner working angles and low stellar flux and is well suited for goals related to spectroscopic characterization \citep{Currie2025,Currie2025b}.

\begin{acknowledgments}

\indent We thank the anonymous referee for a timely and helpful report; Eric Mamajek provided helpful discussions about the spectral type and age of HIP 54515 A.\\
\indent The authors acknowledge the very significant cultural role and reverence that the summit of Maunakea holds within the Hawaiian community.  We are most fortunate to have the opportunity to conduct observations from this mountain.

\indent This research has made use of the Keck Observatory Archive (KOA), which is operated by the W. M. Keck Observatory and the NASA Exoplanet Science Institute (NExScI), under contract with the National Aeronautics and Space Administration.\\
\indent The development of SCExAO was supported by JSPS (Grant-in-Aid for Research \#23340051, \#26220704 \& \#23103002), Astrobiology Center of NINS, Japan, the Mt Cuba Foundation, and the director's contingency fund at Subaru Telescope.  CHARIS was developed under the support by the Grant-in-Aid for Scientific Research on Innovative Areas \#2302.  SCExAO’s adaptive optics loops and high-speed data acquisition are handled by  the CACAO package, which is supported by NSF award 2410616. \\
\indent This work is generously supported by National Science Foundation (NSF) Astronomy and Astrophysics grant \#2408647 and NASA-Keck Strategic Mission Support Proposal.  We are grateful for the continued valuable work and expert guidance of NSF and NASA-Keck personnel, especially in challenging times. M.K. is supported by JSPS KAKENHI (grant No. 24K07108).  A.Z. acknowledges support from ANID -- Millennium Science Initiative Program -- Center Code NCN2024\_001 and Fondecyt Regular grant number 1250249.\\
\indent Finally, this work is dedicated to the memory of Mr. Wallace Ishibashi, Jr.: East Hawai`i community leader, Hawaiian cultural practitioner and Office of Maunakea Management cultural monitor, Royal Order of Kamehameha I member, Department of Hawaiian Homelands commissioner, protector of Maunakea's archaeological and cultural sites, and supporter of the Thirty Meter Telescope as well as the current Maunakea Observatories.  \textit{A hui hou}, Uncle Wally.\\
\end{acknowledgments}


\bibliographystyle{aasjournal}
\bibliography{bibliography}

\appendix
\section{A Recalibrated CHARIS Astrometric Solution\label{astrocal}}

\begin{deluxetable}{lllllllll}[ht]
    \label{obslog_astrocal}
     \tablewidth{0pt}
    \tablecaption{CHARIS Astrometric Calibration Data}
    \tablehead{\colhead{UT Date} & \colhead{Instrument} &{Filter} & {$t_{\rm int}$ (s)}}
    \startdata
    \textbf{M5}\\
    20060703 & Keck/NIRC2 & $K_{\rm p}$ & 108.6\\
    20250315 & SCExAO/CHARIS & $JHK$ & 302.4\\
    20250514 & SCExAO/CHARIS & $JHK$ & 247.8\\
    \textbf{Theta 1 Ori B}\\
    20011120 & Keck/NIRC2 & $NB_{2.108}$ & 200\\
    20041003 & Keck/NIRC2 & $K_{\rm p}$ & 160\\
    20050216 & Keck/NIRC2 & $NB_{2.108}$ & 40\\
    20140903 & Keck/NIRC2 & $K_{\rm p}$ & 19.2\\
    20160118 & Keck/NIRC2 & $NB_{2.108}$ & 225\\
    20191103 & Keck/NIRC2 & $NB_{2.108}$ & 80\\
    20210202 & Keck/NIRC2 & $NB_{2.108}$ & 450\\
    20241122 & SCExAO/CHARIS & $JHK$ & 61.9
    \enddata

    \end{deluxetable}

To affirm the north position angle calibration for our 2022 and 2024 epoch data, we compared the Keck and CHARIS astrometric positions of the HD 1160 B companion \citep{Nielsen2012} and a newly-discovered brown dwarf \citep{ElMorsy2025}, respectively.   Both companions were observed during the same runs as our first two HIP 54515 data sets and thus provide a direct constrain on CHARIS's astrometric calibration.  In both the SCExAO/CHARIS and Keck/NIRC2 HD 1160 data sets, the companion is visible without PSF subtraction techniques; the brown dwarf from \citet{ElMorsy2025} is located at $\rho$ $\sim$ 0\farcs{}3 and detectable at a high SNR with conservative algorithm settings resulting in high throughput.  From these comparisons, CHARIS's astrometric calibration is 
fully consistent with prior work \citep{Currie2018,Currie2022,Chen2023}.  

While the AO3K upgrade installed in May 2024 splits some near-IR light into the wavefront sensor with a dichroic it should not change the CHARIS astrometric calibration. However, we reinvestigate any evidence for a north position angle offset by comparing CHARIS astrometry for M5 and theta 1 Ori B with that obtained from Keck/NIRC2, which has a well-characterized astrometric solution \citep{Yelda2010,Service2016}\footnote{We also considered data from the \textit{Hubble Space Telescope}/Wide Field Camera 3 instrument (Program 13297, PI Giampaolo Piotto) and computed relative astrometry following the methods outlined in \citet{Chen2023}.   Our analysis failed to identify any statistically-signficant north position angle offset, albeit with larger uncertainties than with the NIRC2 data.}.  Table \ref{obslog_astrocal} lists the data. 

Basic reduction of the CHARIS data follows steps performed for HIP 54515 in this work; NIRC2 data reduction follows procedures outlined in \citet{Currie2023a}.   As M5 is a distant globular cluster, we directly compared the CHARIS and NIRC2 astrometry for the 10 stars beyond 0\farcs{}6 from our main guide star ($\alpha$ = 15:18:34.1, $\delta$ = +02:05:00.08) but within the field of view for both instruments.   For theta 1 Ori B, we computed relative astrometry between B1 and the B2, B3, and B4 components for each NIRC2 epoch.  Using these measurements, we predicted the companions' positions in the CHARIS epoch.  

To derive a posterior distribution for the best-estimated CHARIS astrometric solution, we use the affine-invariant Markov chain Monte Carlo (MCMC) ensemble sampler \texttt{emcee} \citep{foreman-mackey+2013} with 50 walkers and 10,000 steps.  We found a pixel scale that is slightly smaller than that from \citet{Currie2018} and \citet{Chen2023}, albeit consistent within errors: 16.10 $\pm$ 0.04.   The north position angle offset is slightly smaller than that from \citet{Currie2018}, albeit within errors, but nearly identical to that from \citet{Chen2023}: -2.02 $\pm$ 0.17.   For this work we adopt a revised pixel scale of 16.10 $\pm$ 0.04 mas pixel$^{-1}$ and a north position angle offset equal to that from \citet{Chen2023} with slightly inflated errors to be conservative: -2.03 $\pm$ 0.27 degrees.

We note that our \texttt{orvara} results are insensitive to whether we adopt this revised calibration or retain the previous one from \citet{Currie2018} or \citet{Chen2023}.   For example, adopting the astrometric calibration from \citet{Currie2018} results in a companion mass posterior of $17.1^{+6.7}_{-4.4}$ $M_{\rm Jup}$ and mass ratio posterior of $0.0085^{+0.0031}_{-0.0022}$.  The mass posterior and mass ratio posterior for the simulation assuming a flat companion mass prior are $18.5^{+9.1}_{-5.0}\,M_{\rm Jup}$ and $0.0093^{+0.0042}_{-0.0025}$.  Minor modifications to the CHARIS astrometric calibration then result in a $\sim$0.1--0.15 $\sigma$ change in HIP 54515 b's mass and a $\sim$0.15--0.2$\sigma$ reduction in its mass ratio.

AO3K+SCExAO and CHARIS are now permanently located behind a Nasmyth platform beamswitcher (NBS) as of 2025 September.  Post NBS, the orientation of the field-of-view for CHARIS is rotated by roughly 90 degrees compared to data presented in this paper and prior CHARIS publications.  Thus, the calibration derived above will no longer be valid: a new, more rigorious astrometric calibration of the upgraded AO3K+SCExAO/CHARIS platform is being carried out in 2025 October (J. Lozi, O. Guyon, in progress).
\end{document}